\documentclass[twocolumn]{aastex63}

\newcommand{\tess}{{TESS}}

\newcommand{\msun}{$M_{\odot}$}
\newcommand{\rsun}{$R_{\odot}$}

\newcommand{\gaia}{{Gaia}}

\received{August 6, 2020}
\revised{September 18, 2020}
\accepted{September 25, 2020}

\shorttitle{NGC 2516 Stellar Spins}
\shortauthors{Healy \& McCullough}

\graphicspath{{./}}

\begin{document}

\title{Stellar Spins in the Open Cluster NGC 2516}

\correspondingauthor{Brian F. Healy}
\email{bfhealy@jhu.edu}

\author[0000-0002-7718-7884]{Brian F. Healy}
\affiliation{Department of Physics and Astronomy\\
Johns Hopkins University \\
3400 North Charles Street \\
Baltimore, MD 21218, USA}

\author[0000-0001-9165-9799]{P.R. McCullough}
\affiliation{Department of Physics and Astronomy\\
Johns Hopkins University \\
3400 North Charles Street \\
Baltimore, MD 21218, USA}
\affiliation{Space Telescope Science Institute \\
3700 San Martin Drive \\
Baltimore, MD 21218, USA}

\begin{abstract}
Measuring the distribution of stellar spin axis orientations in a coeval group of stars probes the physical processes underlying the stars' formation. In this paper, we use spectrophotometric observations of the open cluster NGC 2516 to determine the degree of spin alignment among its stars. We combine \tess\ light curves, ground-based spectroscopy from the Gaia-ESO and GALAH surveys, broadband stellar magnitudes from several surveys, and \gaia\ astrometry to measure 33 stellar inclinations and quantify overall cluster rotation. Our measurements suggest that stellar spins in this cluster are isotropically oriented, while allowing for the possibility that they are moderately aligned. An isotropic distribution of NGC 2516 spins would imply a star-forming environment in which turbulence dominated ordered motion, while a moderately aligned distribution would suggest a more substantial contribution from rotation. We also perform a three-dimensional analysis of the cluster's internal kinematics, finding no significant signatures of overall rotation.
Stemming from this analysis, we identify evidence of cluster contraction, suggesting possible ongoing mass segregation in NGC 2516.

\end{abstract}


\section{Introduction} \label{sec:intro}
Observations and current theories of star formation agree that stars originate from giant molecular clouds (GMCs). Within a GMC, supersonic turbulence shocks the gas and creates clumps of higher density throughout. As each clump's self-gravity overcomes the opposition of magnetic and turbulent gas pressure, high-density cores form within. These cores collapse, heating up and rotating faster through conservation of energy and angular momentum. Eventually a protostar forms in each core's center, surrounded by a nebular disk \citep{shu1987,mckee2007}. Although the theory of star formation has come a long way, there remain many major questions to answer.

One such question is how strongly the imprint of a star-forming cloud's angular momentum is observable on the stars it forms. A straightforward way to answer this question is through the measurement and study of projected stellar spin inclinations: the angle of the star's rotation axis with respect to the observer. For stars that do not experience post-formation tidal torques on their spin axes from sources such as binary companions or close-in giant planets \citep[e.g.][]{hurley2002, ogilvie2014}, stellar inclinations contain information about the initial conditions within their star-forming clump. These initial conditions are set by the proportion of kinetic energy in ordered motion and turbulent flow, as well as the presence of magnetic fields in a clump. Depending on the strength of turbulence, the inclinations of stars formed in a clump will reflect this energy proportion in their relative alignment with each other.  Magnetohydrodynamic numerical simulation shows mutual alignment in the magnetic fields of protostellar cores, resulting from the large-scale conditions of their star-forming region \citep{kuznetsova2020}. Consequently, the spin axes of the resulting stars could be aligned commensurate with the cores' alignment. Other numerical simulations also suggest the possibility of spin alignment among clustered stars if $\gtrsim$ 50\% of their clump's energy is in rotation \citep[][]{corsaro2017,reyraposo2018}.

The orientation of a protostar's spin axis is also connected to the formation of protoplanetary disks. We observe magnetic fields in protostars \citep{crutcher1999}; we observe protoplanetary disks \citep[e.g.][]{boyden2020}; and we theorize that exoplanets originate within such disks \citep{lissauer1993}. Alignment of the magnetic field with the rotation axis of a protostar enhances magnetic diffusion of angular momentum and thereby inhibits disk formation \citep[e.g.][]{mellon2008}. Turbulence may cause misalignment, decreasing angular momentum diffusion and thereby enhancing the formation of protoplanetary disks \citep{joos2012,joos2013}.

The assumption that stellar inclinations are isotropically distributed has a long history in scientific literature \citep[e.g.][]{struve1945}. Since then, there have been numerous studies at different points of stellar evolution, presenting conflicting results: some observations of protoplanetary disks have found evidence of alignment perpendicular to the large-scale magnetic field \citep{tamura1989, vink2005}, but others infer random orientations \citep{menard2004}. A study of planetary nebulae in the galactic bulge found evidence for angular momentum alignment \citep{rees2013}. 

Star clusters in particular offer valuable insight into the physics of star formation because the stars share the same history within a GMC and their physical association is still identifiable now. Previous studies of the spins of open cluster stars have been limited in scope and conflicting in their results: a spectrophotometric study of the Pleiades and Alpha Per clusters found that the distribution of inclinations suggested isotropy in each cluster, but the authors could not rule out anisotropic alignment \citep{jackson2010}. The same collaboration performed another analysis of the Pleiades with new data and once again found results supporting isotropic spins \citep{jackson2018}. \citet{kovacs2018} measured a preferred anisotropic distribution of Sun-like stellar spins in the Praesepe cluster, while \citet{jackson2019} found that isotropy was the most likely scenario for the cluster's M dwarfs. This result is consistent with one of the aforementioned numerical simulations, which only predicted spin alignment in stars $> 0.7$ \msun\ \citep{corsaro2017}. Employing an asteroseismic approach to measure inclinations, \citet{corsaro2017} found significant spin alignment in NGC 6791 and NGC 6811, both of which are several gigayears in age. Emphasizing the discordant results pertaining to spin alignment, however, \citet{mosser2018} used a different approach to analyze the asteroseismic data for these clusters, leading to the conclusion that their spins have no preferential alignment.

It is conceivable that a cluster displaying alignment of stellar spins may also show evidence of overall rotation preserved from its progenitor molecular cloud. Performing a follow-up study of these clusters, \citet{kamann2019} found evidence of bulk rotation in NGC 6791, with an orientation consistent with the average inclination of its stars as measured by \citet{corsaro2017}. The agreement of these results suggested a possible connection between the overall rotation of a cluster and its stars' spins, even after billions of years of evolution. 

We can measure projected stellar inclinations by comparing the spectroscopically measured projected rotational velocity of the star $v\sin i$ to the velocity derived from the stellar circumference $2\pi R$ divided by the photometric period of rotation $P$:

\begin{equation}
    v\sin i = \frac{2\pi R}{P}\sin i,
    \label{eq:spectrophot}
\end{equation}
Measuring $\sin i$, the sine of the inclination to the observer's line of sight (LOS) of the star's spin axis, requires a combination of measurements of the three observational quantities obtained from multiple sources.
With the arrival of precise data from \tess\ \citep{ricker2015} and \gaia\ \citep{gaia2016} combined with public ground-based surveys, we are in a newfound position to quantitatively measure stellar spin alignment in many open clusters.

In this paper, we use the above spectrophotometric method to determine the projected inclinations of 33 stars in the open cluster NGC 2516 ($l = 273^{\circ}.861$, $b = -15^{\circ}.873$, distance $\sim409$ pc \citep{cantatgaudin2018}, age $\sim$140 Myr \citep[e.g.][]{meynet1993}, mass $\sim 1240 - 1560$ \msun\ \citealp{jeffries2001}). We access ground-based surveys for $v\sin i$ values, measure rotation periods using \tess\ light curves, and determine precise stellar radii through spectral energy distribution (SED) fitting to a host of stellar magnitudes paired with a \gaia\ parallax. We use Bayesian statistics to compute inclinations and their uncertainties given these three parameters. We also analyze cluster kinematics using \gaia\ proper motions and ground-based LOS velocity measurements.

The rest of the paper proceeds as follows: Sec.\ \ref{sec:obs} describes the sources of data we use to measure projected inclinations. Sec.\ \ref{sec:analysis} presents the methods of analysis we employ to make the measurements. In Sec.\ \ref{sec:results} we report the results for this study, including 150 rotation periods for cluster members and 33 inclinations. We discuss these results in Sec.\ \ref{sec:discussion} and conclude with Sec.\ \ref{sec:conclusion}.

\section{Data Sources and Reduction}
\label{sec:obs}

\subsection{Cluster membership from Gaia}
\label{subsec:membership}
To identify the cluster stars appropriate for this study, we referenced an NGC 2516 membership table from \citet{cantatgaudin2018} based on \gaia\ astrometry. From this table, we selected 650 stars with a membership probability greater than 0.68 and a measured $G_{\rm BP} - G_{\rm RP}$ color. We optimized the membership probability cutoff to balance the number of stars with high confidence of membership. We also adopted the mean cluster parameters for NGC 2516 from \citet[][]{cantatgaudin2018}, including position on the sky and proper motion. We accessed the Gaia DR2 archive \citep{gaia2018} for parallax measurements and astrometric goodness-of-fit parameters.

\subsection{Gaia-ESO and GALAH spectroscopy}
To supply $v\sin i$ measurements for our analysis and further refine cluster membership with radial velocities, we used data from two Southern Hemisphere spectroscopic surveys: Gaia-ESO  \citep[GES,][]{gilmore2012} and GALAH \citep{desilva2015}. The GALAH DR2 data are publicly available \citep{buder2018}, and we accessed published GES data for NGC 2516 from Table 1 of \citet{jackson2016}. We identified 288 stars observed by GES meeting the membership probability cutoff in Sec.\ \ref{subsec:membership}. In the GALAH survey, we found 19 such stars. There were three cluster members observed by both surveys: we discarded one star owing to inconsistent radial velocities and kept the more precise GALAH data for the other two.

\subsection{TESS full-frame images}
NGC 2516 is close to the southern continuous viewing zone of \tess, so its stars received as many as seven sectors of observations. The maximum possible coverage consisted of sectors 1, 4, 7, 8, 9, 10, and 11, while some stars received fewer sectors of coverage. For stars meeting the membership criteria of Sec.\ \ref{subsec:membership}, we accessed light curves from the PATHOS project \citep{nardiello2019_pathos1,nardiello2020_pathos2} and the Cluster Difference Imaging Photometric Survey \citep[CDIPS;][]{bouma2019}. To supplement these data products, we also used the \texttt{eleanor} software and postcards \citep{feinstein2019} to generate light curves with the \texttt{crowded\_field} keyword to use smaller apertures for each star. To remove systematic trends without suppressing rotational signals, we applied the software's principal component analysis (PCA) correction, using only the first three cotrending basis vectors for each camera. For the CDIPS light curves, we used the \texttt{PCA1} data (1 pixel aperture radius).

For PATHOS, we used the \tess\ magnitude-dependent results from Sec.\ 2.6 of \citet[][]{nardiello2020_pathos2} to determine whether we used the light curve from point-spread function fitting or one of the photometric apertures with radius of 1-4 pixels. We also followed that paper's method of ``cleaning'' light curves by excluding from the analysis all points with a low-quality flag (\texttt{DQUALITY} $> 0$), all flux values discrepant with the median flux by more than 3.5$\sigma$, and any point with a measured sky background more than 5$\sigma$ greater than the median value.

\subsection{Public ground-based photometry} \label{subsec:publicphot}
For each cluster star found to have a rotation signature along with a $v\sin i$ measurement, we created an SED for fitting to determine the radius. To build the SED, we used stellar magnitudes from the ALLWISE \citep{wise_wright2010,cutri2013}, Two Micron All Sky Survey \citep[2MASS;][]{skrutskie2006}, Hipparcos/Tycho-2 \citep{perryman1997,hoeg2000}, and APASS \citep{henden2016} databases. For some fainter stars lacking optical magnitudes from Tycho or APASS, we obtained Johnson $B$ and $V$ magnitudes from the \tess\ Input Catalog v8 \citep{stassun2018}.

\begin{figure*}
    \centering
    \includegraphics[scale=0.5]{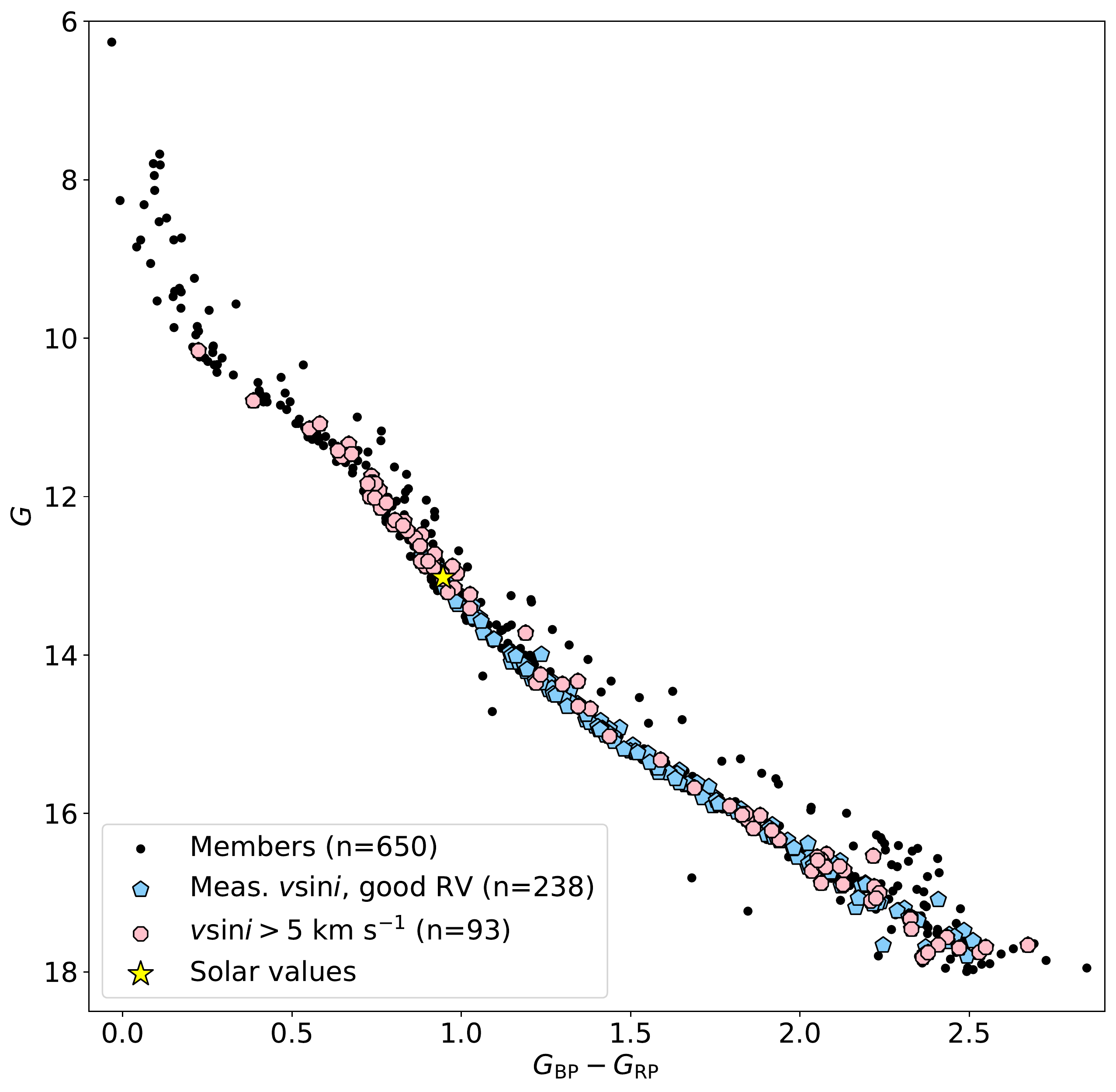}
    \caption{NGC 2516 color-magnitude diagram in \gaia\ apparent magnitudes. Black points represent the 650 high-probability members according to \citet{cantatgaudin2018}. Light-blue points denote the stars that fall below the equal-mass binary main sequence (see Sec.\ \ref{subsec:selection}), have measured $v\sin i$ from either the Gaia-ESO or GALAH surveys, and have radial velocities consistent with that of NGC 2516. Pink points show a subset of that sample with $v\sin i$ greater than a 5 km s$^{-1}$ threshold. The star symbol indicates reddened and solar values at the cluster distance.}
    \label{fig:cmd_singlestars}
\end{figure*}

\section{Data Analysis} \label{sec:analysis}

\subsection{Removing equal-mass binaries}
\label{subsec:selection}
Binary systems of near-equal mass are not suitable for this study, since their component stars' comparable luminosity imprints two sets of similar lines on the spectrum, significantly increases the flux of the SED, and confounds the assignment of rotation periods. Thus, we excluded stars on the equal-mass binary main sequence for NGC 2516, which can be differentiated from the single-star main sequence using \gaia\ photometry.

We divided the cluster's color-magnitude diagram (Fig.\ \ref{fig:cmd_singlestars}) into 30 bins in \gaia\ $G_{\rm BP} - G_{\rm RP}$ color and computed the median $G$ magnitude in each bin. We interpolated in $G$ magnitude between these bins and calculated the residuals of each star's magnitude when compared to the interpolation. Visually identifying a central distribution in the residuals representing the single-star main sequence, we performed another binning and interpolation on this subset of stars. We established a magnitude cutoff of $3\sigma$ brighter than the single-star main sequence, above which we discarded stars as likely multistar systems. We also removed stars with \gaia\ astrometric excess noise significance greater than $3$ (indicating likely binary systems) and eliminated known spectroscopic binaries from this subset using the SIMBAD database \citep{wenger2000}. This culling resulted in 535 remaining stars.

\subsection{Incorporating spectroscopy}
\label{subsec:rvs}
We cross-matched using \gaia\ source IDs to select the 252 out of the 535 single-star cluster members that had either Gaia-ESO or GALAH spectroscopy available. We identified a clustering of radial velocity (RV) measurements at $23.8 \pm 1.1$ km s$^{-1}$ and excluded 14 stars with RVs more than $3\sigma$ discrepant from this value, leaving 238 stars in this sample.

Stars that are slowly rotating or viewed pole-on (or both) will display low levels of Doppler broadening. For sufficiently small values of $v\sin i$, the rotational broadening cannot be disentangled reliably from other velocity-broadening mechanisms, including turbulent motion on the scale of $\sim$ 1 km s$^{-1}$. Since biased $v\sin i$ values can lead to systematic inclination errors \citep[e.g.][]{kamiaka2018}, we established a minimum acceptable $v\sin i$ threshold of 5 km s$^{-1}$, in accordance with the threshold recommended by \citet[][]{jackson2016}.

We determined that 93 stars in NGC 2516 met all of the above criteria. Fig.\ \ref{fig:cmd_singlestars} illustrates the subsamples described above. We also plot $G$ and $G_{\rm BP} - G_{\rm RP}$ values for the Sun based on \gaia\ solar magnitudes \citep[$M_{G,\sun} = 4.67$ and ${(G_{\rm BP} - G_{\rm RP})}_{\sun} = 0.82$,][]{casagrande2018}, a cluster distance measurement \citep[409 pc;][]{cantatgaudin2018}, a reddening estimate \citep[$E(B-V) = 0.12,$][]{jackson2016}, and a conversion of this estimate to extinction and reddening in \gaia\ passbands \citep{wang2019} under the assumption that $R_{V} = 3.1$.

\subsection{Rotation periods}
\begin{figure*}
    \centering
    \includegraphics[scale=0.5]{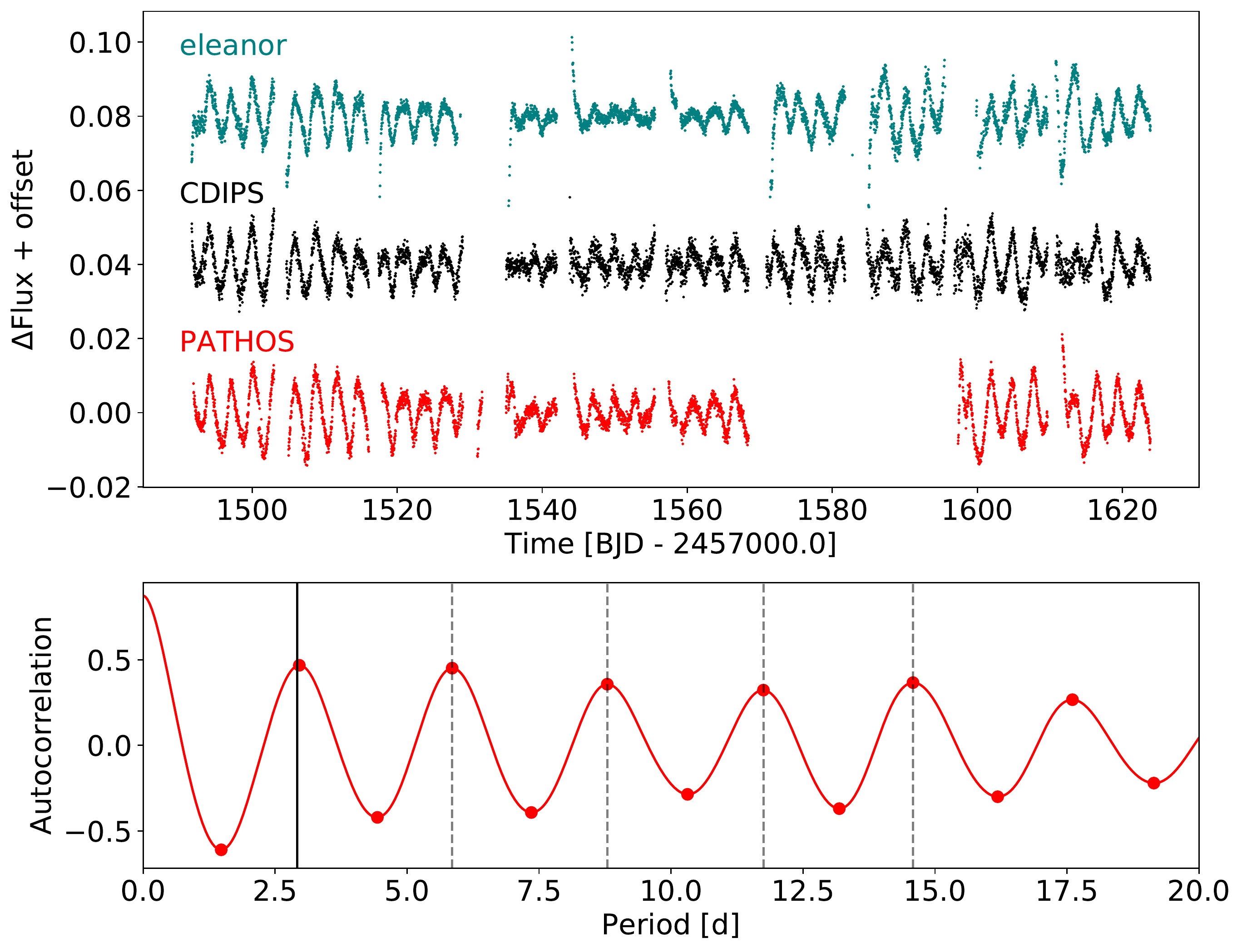}
    \caption{{Top:} Sectors 7-11, \tess\ light curves from eleanor, CDIPS, and PATHOS for Gaia DR2 5290728834785867264, a $T$ = 12.37, $G$ = 12.84 Sun-like star. {Bottom:} smoothed autocorrelation function of the PATHOS light curve for this star. The first peak (solid line) is located at the rotation period, found to be $2.921 \pm 0.048$ days. The agreement between the predicted locations of the next four harmonic peaks (dashed lines) and their actual locations indicates a precise period determination. We do not use more than five peaks in this analysis to avoid a loss of period accuracy over long lag times.}
    \label{fig:lc_acf}
\end{figure*}

For all 535 likely cluster members on the single-star main sequence, we analyzed the \tess\ light curves from PATHOS, CDIPS, and \texttt{eleanor} to measure stellar rotation periods. Our primary method of determining each period was using the light curve's autocorrelation function (ACF), following the technique of \citet{mcquillan2013,mcquillan2014}. The smoothed ACF of a periodic light curve yields another periodic function whose peaks represent integer multiples of the period (e.g.\ Fig \ref{fig:lc_acf}). Nominally, the location of the first peak in the ACF corresponds to the rotation period, and all subsequent peaks represent harmonics with decreasing amplitude. In cases where a star has multiple spot groups at different longitudes, the star's brightness may display complex modulation for a single revolution of the star, creating a potential to underestimate the period. Typically spot groups are not identical, and in that case, the second ACF peak will have a greater amplitude than the first, indicating that the true rotation period is the longer of the two peaks' periods. 

To prepare a light curve for autocorrelation, we normalized it and subtracted unity from each point. We mapped the flux values onto a uniformly spaced array of time points, filling gaps in the data with values of zero. We computed an ACF for each star using a routine from the \texttt{emcee} code \citep{foremanmackey2013}, smoothing the ACF to enable reliable peak and trough detection. We calculated a relative height for each peak based on its separation from neighboring troughs.

We assigned periods and uncertainties to each light curve in one of two ways. For light curves for which the periodicity was subtle, the ACF showed a single peak. We adopted the abscissa coordinate of the peak as the rotation period, and we estimated the uncertainty by measuring the half-width at half-maximum of the peak. For light curves with higher signal-to-noise ratios where sinusoidal modulation is clearly visible, the ACF shows many peaks located at integer multiples of the period. We incorporated the information given by these multiple peaks into a more precise period measurement by performing a linear fit to the peak positions as a function of peak number. We included the coordinate (0, 0) and as many as five peaks in this fit, limiting the number of peaks to avoid a loss of accuracy in the period across long lag times \citep{mcquillan2013}. We took the slope of the line as the period and defined the uncertainty using the  standard deviation of each peak's position relative to harmonics of the period.

This algorithm does not evaluate whether or not a periodic signal is due to stellar rotation. Therefore, we visually inspected a report for every star that included the full light curve, ACF, Lomb--Scargle periodogram \citep[e.g.][]{nielsen2013} calculated with \texttt{lightkurve} \citep{lightkurve2018}, \tess\ pixel cutout, list of nearby stars, and period provided by the algorithm. 
While the periodogram method struggles with the aforementioned scenario of multiple spot groups, it works well in cases of rapid rotation ($P < 1\ {\rm day}$), where the smoothed ACF may blend peaks and report an overestimated period \citep{mcquillan2013}. We used periodograms to identify rapid rotator candidates, and we then decreased the amount of ACF smoothing in a second analysis to detect the true period. We also consulted gyrochronological predictions \citep[e.g.][]{barnes2003} for expected rotation periods as a function of color for stars of a known age. This information aided our verification of blended cluster members with different colors.

Based on the elements of each period report, we rejected signals that showed no modulation, inconsistent periodicity between sectors, binary eclipses, and brightness variations induced by scattered light in the \tess\ field of view. We also eliminated periodic astrophysical signals not caused by rotation: signals due to binary eclipses and asteroseismic oscillations differed respectively in shape and amplitude/frequency compared to starspot modulation, allowing for their identification and exclusion. Finally, we used Gaia DR2 to identify stars blended within one \tess\ pixel. In cases of blended stars of comparable color and magnitude, we were unable to verify the source of rotational modulation. Thus, we did not assign a rotation period for these signals despite their likely rotational origin.

\subsection{Stellar radii}
\label{subsec:radii}
We further culled the sample of stars with $v\sin i$ $>$ 5 km s$^{-1}$ and measured rotation periods by limiting the acceptable effective temperature range to between 4000 and 10,000 K. This selection kept main-sequence stars bright enough to have high signal-to-noise ratio magnitude measurements, while it excluded faint M dwarfs on the red end and evolving stars on the blue end. After manually removing a few additional stars whose \gaia\ apparent magnitude and colors placed them off of the cluster's single-star main sequence, we determined the radii of this 43-star sample via SED fitting \citep[e.g.][]{stassun2017} using the broadband stellar magnitudes identified in Section \ref{subsec:publicphot}.

We used EXOFASTv2 \citep{eastman2019} to simultaneously fit an SED and MIST isochrones \citep{choi2016,dotter2016} to each star's magnitude data, establishing several Gaussian priors for each Markov Chain Monte Carlo (MCMC) run, attempting to avoid overconstraining them by using conservative uncertainties. We set individual parallax priors from \gaia\ DR2, increasing each parallax by $30$ $\mu$as to correct for the systematic offset shown by \citet{lindegren2018}. We also increased the parallax uncertainty by using the formula in \citet{lindegren2018p}, and by propagating the offset as an additional error. We also set individual effective temperature priors with $5\sigma$ uncertainties based on Gaia-ESO or GALAH spectroscopy.

We also established Gaussian priors based on measurements for the entire cluster: we set a mean $A_{V}$ extinction prior of $0.22 \pm 0.4$ to be consistent with estimates from multiple sources \citep{jackson2016, bossini2019}. We used two different age priors: one based on the consensus age of $\sim 140 $ Myr \citep[e.g.][]{meynet1993}, and the other using a more recent estimate of $\sim 250$ Myr \citep{bossini2019}. Each age prior had an uncertainty of $\sim 20\%$. We found that the age discrepancy does not produce significantly different results in the radii. We also constrained each star's [Fe/H] to the cluster's mean and uncertainty ($-0.2 \pm 0.4$), informed by the GES and GALAH spectroscopic surveys and existing literature values \citep[e.g.][]{jeffries2001, sung2002}.

\begin{figure*}
    \centering
    \includegraphics[scale=1]{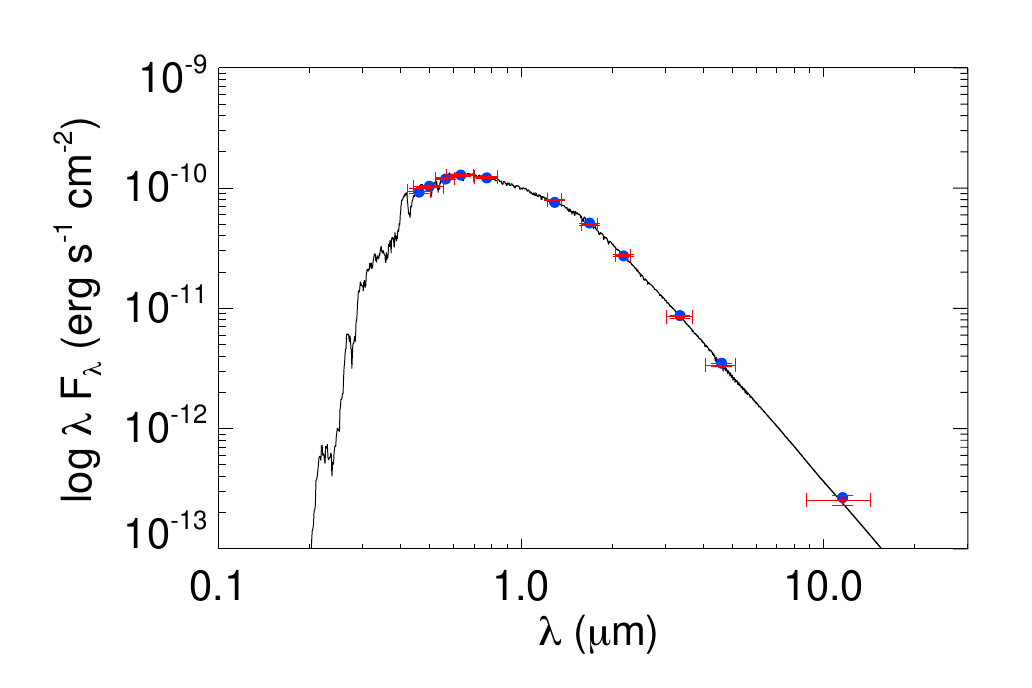}
    \includegraphics[scale=1]{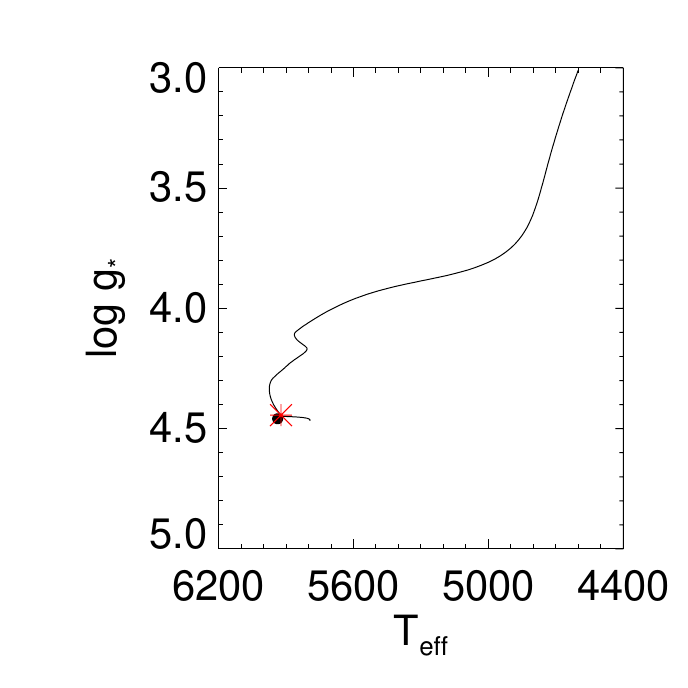}
    \caption{{Left:} EXOFASTv2 SED fit for Gaia DR2 5290728834785867264. The measured APASS, 2MASS, and ALLWISE magnitudes and uncertainties are shown in red. The black curve is the best-fitting model SED, and the blue points are the mean model flux in each passband. {Right:} MIST isochrone fit for Gaia DR2 5290728834785867264. The black dot is the best fit for the star, while the red asterisk shows the nearest model value on the isochrone (black curve).}
    \label{fig:exofast_sed}
\end{figure*}

Along with these priors, we provided unconstrained starting values for the stellar radius and mass from the TESS Input Catalog \citep{stassun2018}. Finally, we set the equivalent evolutionary point (EEP, used by the MIST isochrone) for each star based on a short preliminary fit. Fig.\ \ref{fig:exofast_sed} shows an example of the outputs of SED and MIST isochrone fitting for Gaia DR2 52907288347858672, one of the stars for which we measured $\sin i$.

To ensure that all results from the MCMC runs were well mixed, we required the Gelman--Rubin statistic to be less than 1.01. We discarded stars that did not satisfy the mixing requirement at the end of their run, and we removed others that showed an unsatisfactory isochrone fit, perhaps due to a companion in the system. With all cuts to the sample complete, 33 stars met all the criteria to determine the sine of the inclination.

\subsection{Determination of $sini$}
\label{subsec:determining_sini}
From our rotation period and radius measurements, we calculated each star's rotation speed using $v = \frac{2 \pi R}{P}$. This calculation assumes that the effect of stellar differential rotation is negligible (see Sec.\ \ref{subsec:discussion_periods}). We determined the uncertainty in $v$ by propagating the errors in $R$ and $P$. Our determination of $\sin i$ involved a comparison between the calculated $v$ and measured $v\sin i$, two correlated parameters. Therefore, we used the procedure of \citet{masuda2020} to perform Bayesian inference and determine the posterior probability distribution (PPD) of $\sin i$ for each star given the data $D$.

We used each star's $v$ and $v\sin i$ measurements and uncertainties to define likelihood functions $\mathcal{L}_{v}(v)$ and $\mathcal{L}_{v\sin i}(v\sin i)$. We used a Gaussian distribution to describe $\mathcal{L}_{v}(v)$ and Student's $t$-distribution for $\mathcal{L}_{v\sin i}(v\sin i)$, in accordance with the results of \citet{jackson2015}. As in \citeauthor[][]{masuda2020}, we set uniform priors $\mathcal{P}_{v}(v)$ and $\mathcal{P}_{\sin i}(\sin i)$ for the rotation speed and $\sin i$, respectively. To properly account for the dependence between $v$ and $v\sin i$, we applied Bayes's theorem to set up an integral calculating $p(\sin i\ |\ D)$, the PPD for $\sin i$:
\begin{equation}
\label{eq:bayesprob}
    p(\sin i\ |\ D) \propto \mathcal{P}_{\sin i}(\sin i) \int \mathcal{L}_{v}(v) \mathcal{L}_{v\sin i}(v\sin i) \mathcal{P}_{v}(v) {\rm d}v
\end{equation}
The low dimensionality of this problem enabled us to directly compute the integral to obtain the PPD of $\sin i$ for each star. From these distributions, we adopted the median value as the sine of the inclination, with the 16th and 84th percentiles providing the uncertainties. 

\subsection{Analysis of inclination distribution}
\label{subsec:distribution}
To determine the degree of alignment among the 33 measured $\sin i$ values for NGC 2516, we compared the data with a model based on \citet{jackson2010}. This model imagines stellar rotational poles randomly filling a bipolar cone defined by two parameters: the mean inclination $\alpha$ and the half-angle of the inclinations' spread, $\lambda$. The angle $\alpha$ ranges from $0^{\circ}$ (pole-on orientation) to $90^{\circ}$ (edge-on orientation). The angle $\lambda$ shares the same range, with $0^{\circ}$ corresponding to complete alignment and $90^{\circ}$ representing complete isotropy. One can calculate the cumulative distribution function (CDF) of projected inclinations for any pair of $\alpha$ and $\lambda$, and compare to the observed CDF.

The cone model also includes a fitted parameter to determine the threshold $\sin i$ value for the model that represents the most pole-on detectable inclination. A completely pole-on star will display no rotational broadening of spectral lines, and any periodic modulation in its light curve will also be minimal. Accounting for this threshold in the model makes it more realistic when compared to our $\sin i$ measurements.

In addition to the $\sin i$ threshold, we modified the model distribution function to allow for the typical measurement uncertainties in $P$, $R$ and $v\sin i$ that can result in values of $\sin i > 1$. These seemingly nonphysical cases are important to keep in the sample: they represent inclinations that are nearly edge-on, and excluding them artificially changes the distribution of spins for the whole cluster.

We used a least-squares method to determine the best model fit to the cumulative distribution of the data given all possible combinations of the $\sin i$ threshold (in intervals of 0.05) and $\alpha$ and $\lambda$ (in $1^{\circ}$ intervals).

To provide another metric evaluating the degree of spin alignment, we calculated the ``alignment coefficient'' $A$ of our measurements, as shown in  \citet{corsaro2017}:
\begin{equation}
\label{eq:aligncoeff}
    A = \frac{1}{N} \sum_{j=1}^{N} \cos^2 i_{j}.
\end{equation}
As $N \to \infty$, the alignment coefficient for a completely isotropic distribution converges to $1/3$. A perfectly aligned distribution yields $A = 1$.

\subsection{Cluster rotation}
\label{subsec:clusterot}
Motivated by the work of \citet[][]{kamann2019}, we consulted \gaia\ proper-motion measurements for NGC 2516 combined with public RVs to analyze the three-dimensional motion of cluster stars, searching for any detectable rotation.

After mapping each star's measurements from celestial coordinates to Cartesian space using Eq.\ 2 of \citet[][]{gaiaclusters2018}, we determined each star's radial distance from the center of NGC 2516 using the cluster's position from \citet[][]{cantatgaudin2018}. We also calculated each star's position angle $\theta_{j}$. We then put stars into bins based on distance from cluster center. We placed the 535 likely members on the single-star main sequence into seven bins. The first six of these bins contained 80 stars, while the final bin contained 54.

We referenced the same spectroscopic data as our inclination analysis for 238 LOS velocities that measure the third dimension of stellar motion. Using Eq.\ 6 of \citet[][]{vandeven2006}, we calculated the expected contribution of perspective rotation to the LOS velocities and subtracted these values from the data.

We subtracted the mean proper motion of NGC 2516 from each of its member stars' measurements. Adopting
a cluster distance of $408.9 \pm 0.1$ pc from \citet[][]{cantatgaudin2018}, we converted angular velocities to linear velocities in units of km s$^{-1}$, propagating astrometric errors as independent uncertainties. We transformed each star's motion in R.A.\ and decl.\ into polar coordinates to separate the tangential ($\theta$) and radial ($r$) components of stellar motion using Eq.\ 10 of \citet{vanleeuwen2000}.

For each of the three directions of motion, we analyzed stellar velocities following the approach of \citet[][]{kamann2019}. Our goals were to (1) determine the radial dependence of the mean velocity $v_{0}$ and dispersion $\sigma$ in the plane of the sky, (2) search for trends in these quantities with radial distance, and (3) estimate the cluster's mean position angle and rotation rate in the LOS direction. We maximized the likelihood function
\begin{equation}
    \ln \mathcal{L} = \sum_{j=1}^{N} - \ln \Big( 2 \pi \sqrt{\sigma^2 + \epsilon_{j}^2} \Big) - \frac{(v_{i} - v_{0})^2}{2(\sigma^2 + \epsilon_{j}^2)},
\label{eq:rot_likelihood}
\end{equation}
where $v_{i}$ is a single star's velocity, coming from either LOS velocity measurements or proper motions, and $\epsilon_{i}$ quantifies the measurement uncertainty. The definition of $v_{0}$ varied depending on the component being studied: in the directions along the plane of the sky, $v_{0}$ is simply $v_{\rm sys}$, the systemic velocity in that direction. In the LOS direction, $v_{0} = v_{\rm sys}\ +\ v_{\rm rot} \sin (\theta_{j} - \theta_{\rm c})$, where $v_{\rm rot}$ is the rotation speed about an axis perpendicular to the LOS, $\theta_{j}$ is an individual star's position angle, and $\theta_{\rm c}$ is the position angle of the cluster's rotation axis.

To explore the parameter space of our results, We performed MCMC runs using \texttt{emcee}, first for the entire cluster and then for each radial bin (when applicable). For the entire-cluster runs, we used uniform priors in all parameters, limiting the cluster position angle to $0^{\circ} \le \theta_{c} < 360^{\circ}$ and allowing $v_{\rm rot}$ to take only positive values. Since the uncertainties in Gaia-ESO RVs represent a $t$-distribution rather than a Gaussian, we multiplied them by 1.09 to approximate 68\% confidence intervals \citep{jackson2015}. For each run, we used 100 chains and 5000 steps, discarding a burn-in of 500 steps. We adopted the median posterior value to be our measurement for each parameter. We used the 16th and 84th percentiles of each posterior distribution to quantify our measurement uncertainties.

\section{Results} \label{sec:results}
\subsection{Rotation periods}

\begin{figure*}
    \centering
    \includegraphics[scale=0.4]{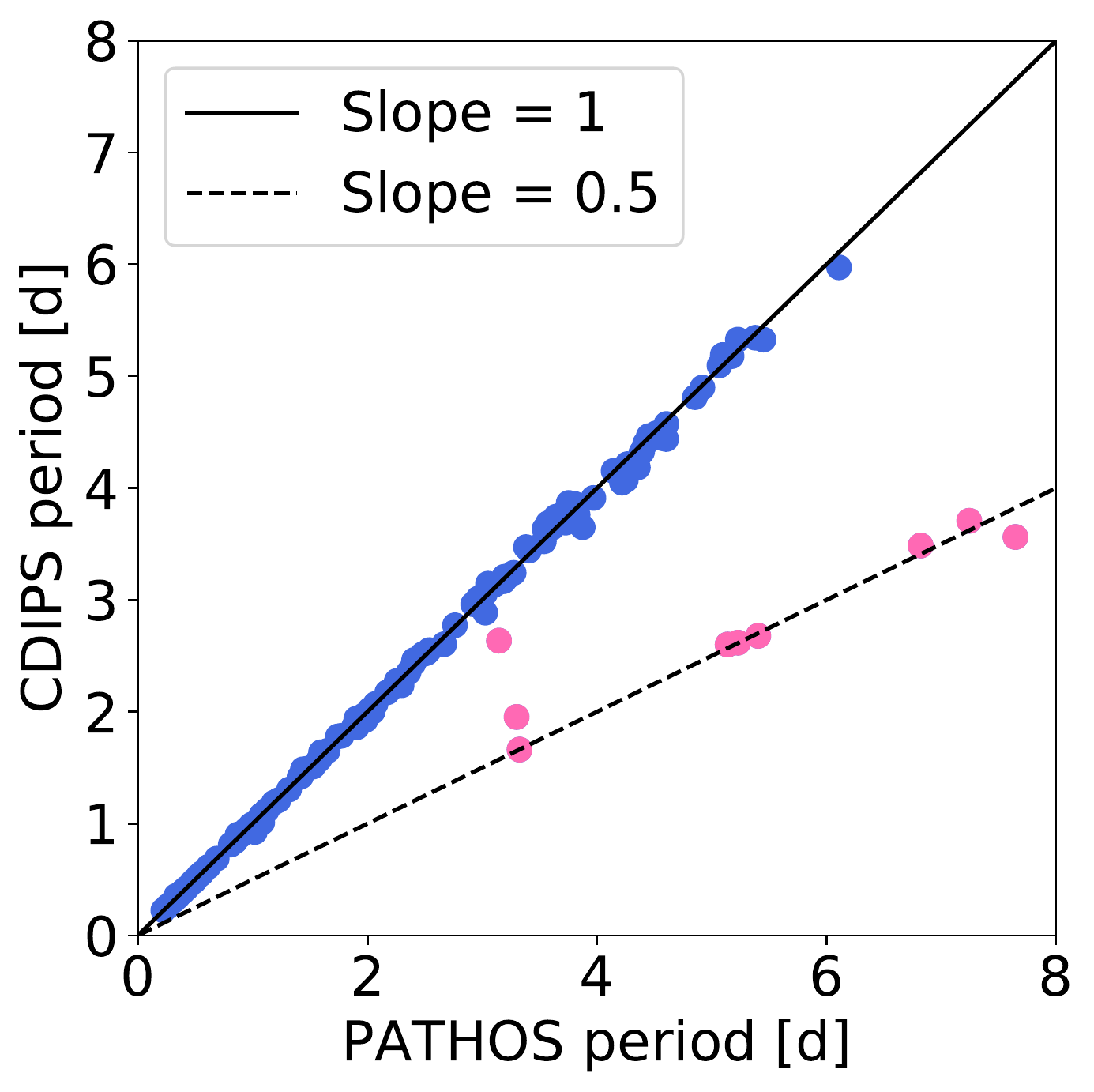}
    \includegraphics[scale=0.4]{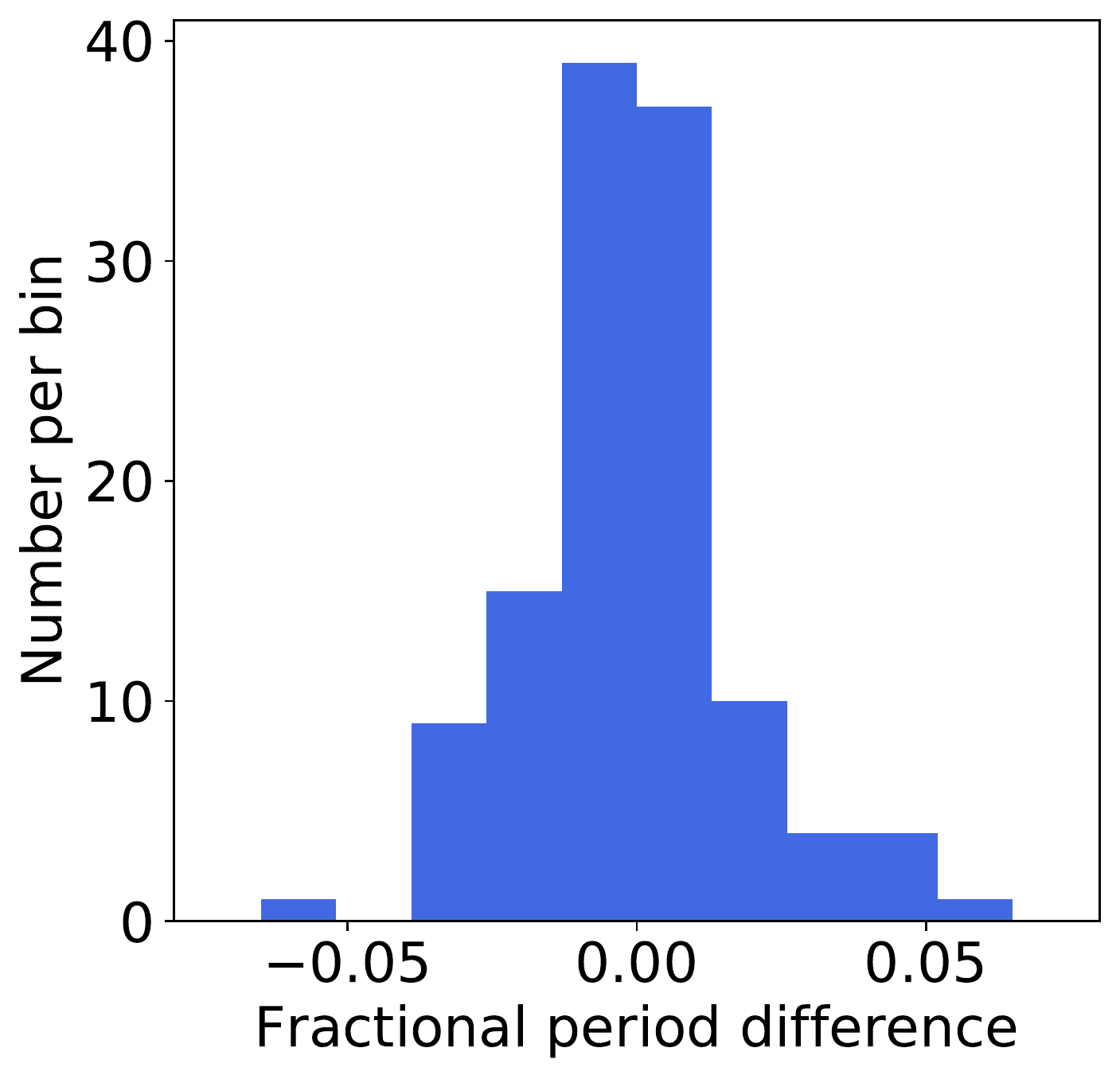}
    \caption{{Left}: Comparison of 131 PATHOS and CDIPS periods, with pink points showing nine measurements discrepant by more than 10\%. Most of the discrepant periods were a factor of $\sim 2$ greater in PATHOS compared with CDIPS. {Right:} fractional difference between the 122 consistent periods, showing most measurements within a few percent of each other.}
    \label{fig:periodcompare_PC}
\end{figure*}

\begin{figure*}
    \centering
    \includegraphics[scale=0.4]{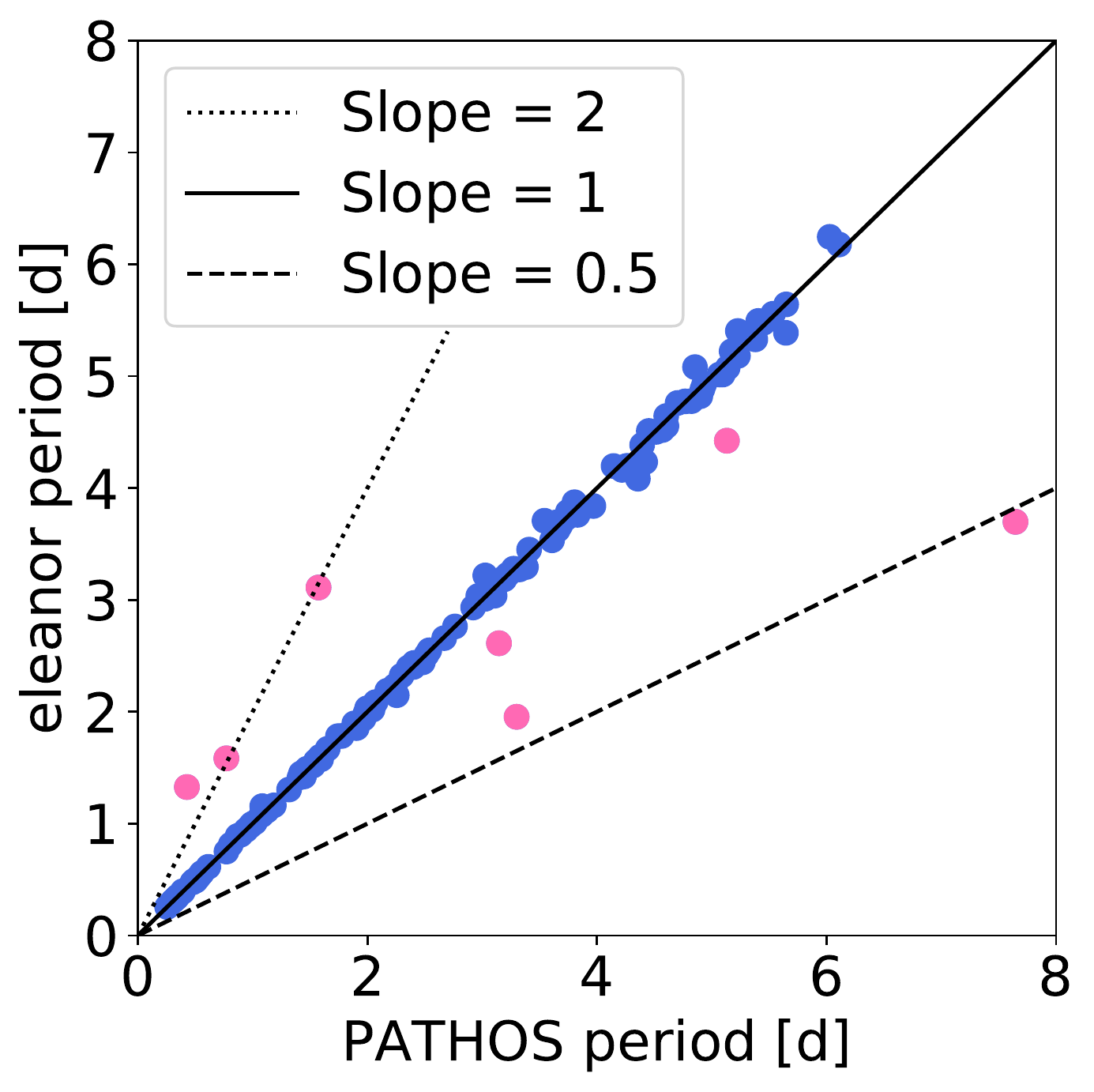}
    \includegraphics[scale=0.4]{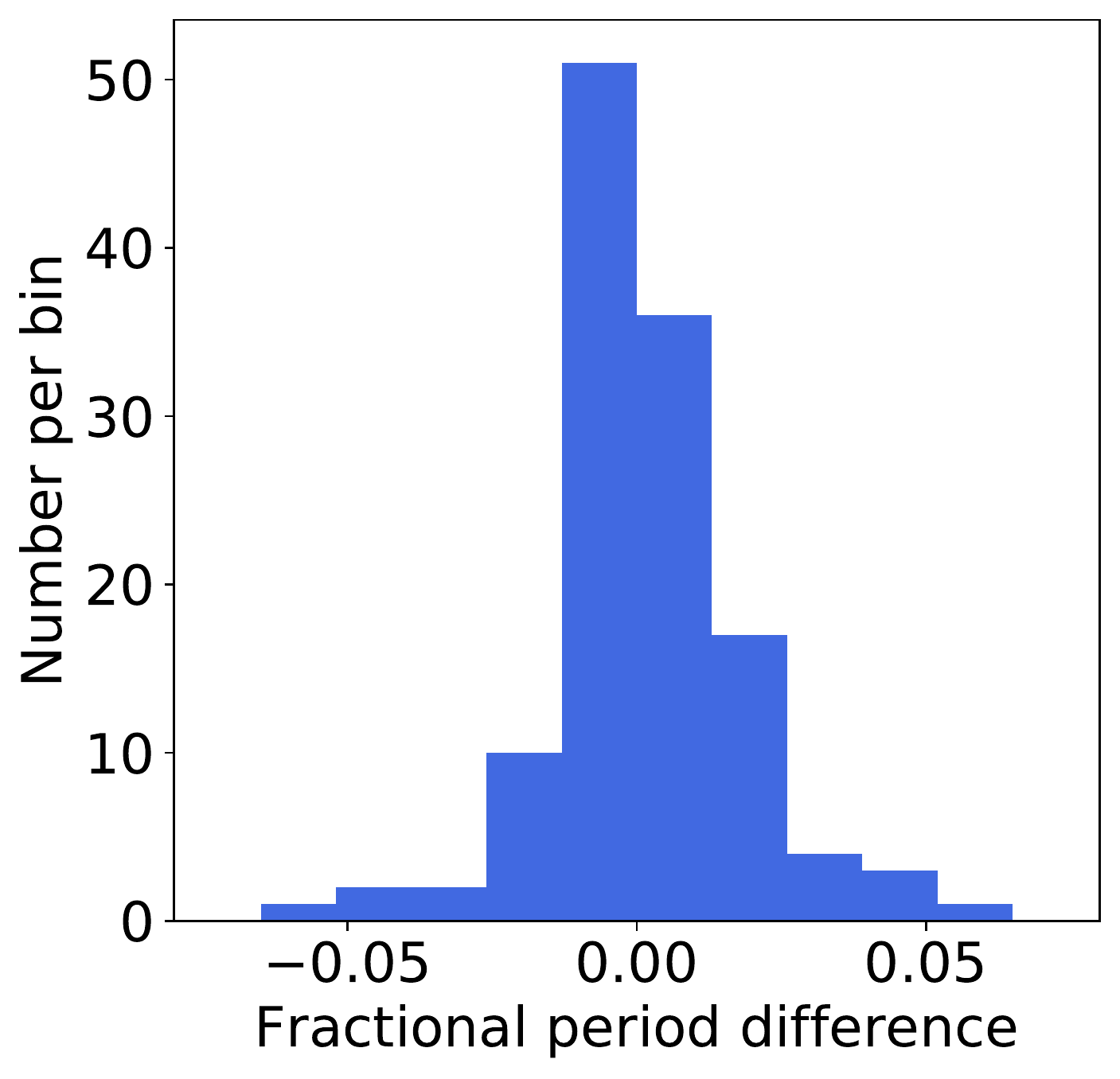}
    \caption{{Left}: Comparison of 135 PATHOS and \texttt{eleanor} periods, with pink points showing seven measurements discrepant by more than 10\%. {Right:} fractional difference between the 128 consistent periods, showing most measurements within a few percent of each other.}
    \label{fig:periodcompare_PE}
\end{figure*}

\begin{figure*}
    \centering
    \includegraphics[scale=0.5]{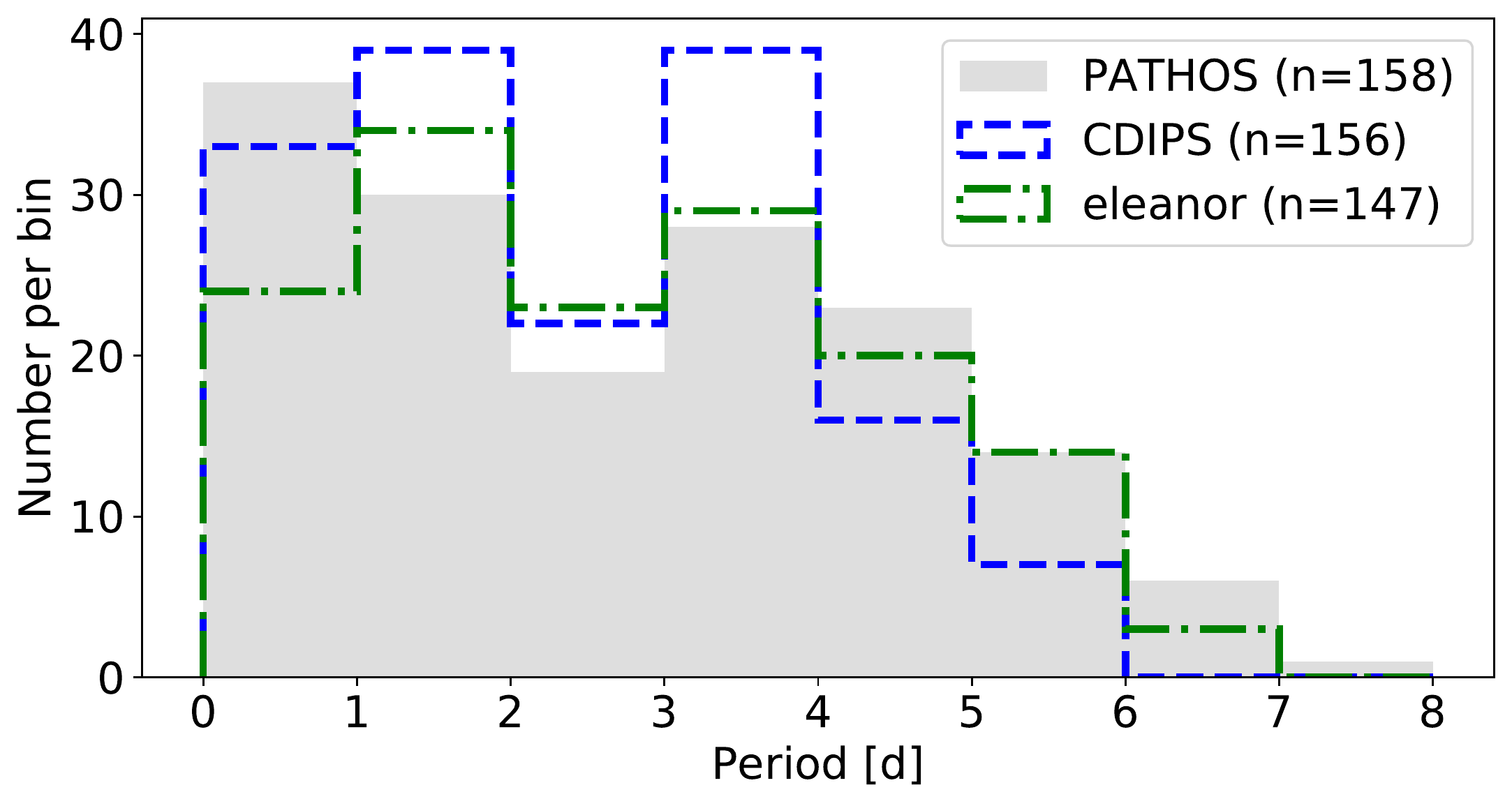}
    \caption{Histogram of rotation periods measured using PATHOS (gray), CDIPS (blue), and \texttt{eleanor} (green) \tess\ light curves. We detected the most periods with PATHOS, and this data set also revealed the greatest number of periods $<1$ day and $>4$ days. CDIPS identified more periods between 1 and 4 days, but fewer than PATHOS at periods $>4$ days.}
    \label{fig:periodcomapre_hist}
\end{figure*}

Out of the 535 likely NGC 2516 members on the single-star main sequence, we identified 171 with a rotation signature using PATHOS light curves, 167 using CDIPS, and 147 using \texttt{eleanor}. Since PATHOS revealed the most rotation periods of the three sources, we adopted these measurements as our default set of periods. To report only the most confident periods, we removed 13 stars from the PATHOS sample that had measurements conflicting with either of the two other data sets by more than 10\%. The left panels of Figs.\ \ref{fig:periodcompare_PC} and \ref{fig:periodcompare_PE} illustrate the comparison between data sets, while the right panels show histograms of the remaining period discrepancies after the selection. Fig.\ \ref{fig:periodcomapre_hist} shows the distribution of the resulting periods for each data set after the removal of conflicting measurements. Combining 137 PATHOS periods consistent with at least one other data source with 21 periods that lacked a comparable CDIPS or \texttt{eleanor} measurement, we report periods for 158 NGC 2516 stars ($\sim 30\%$ of the sample) in Table \ref{tab:periods}.

\begin{figure*}
    \centering
    \includegraphics[scale=0.5]{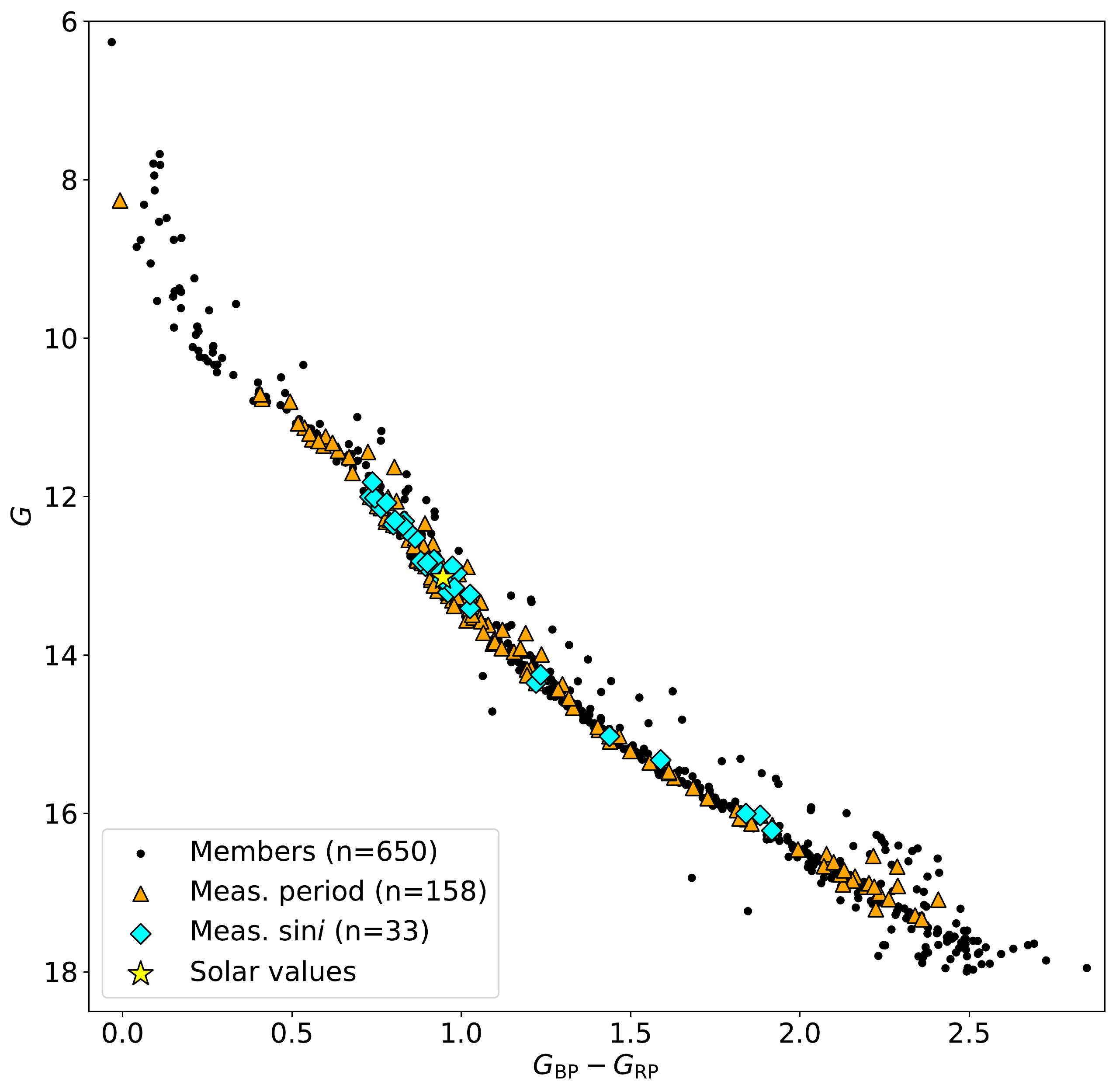}
    \caption{Color--magnitude diagram showing 650 NGC 2516 members (black), 158 members with a measured rotation period (orange), and 33 with measured $\sin i$ (cyan). Most of the latter stars are Sun-like, as illustrated by the star symbol indicating reddened and extincted solar values.}
    \label{fig:cmd_results}
\end{figure*}

\begin{figure*}
    \centering
    \includegraphics[scale=0.5]{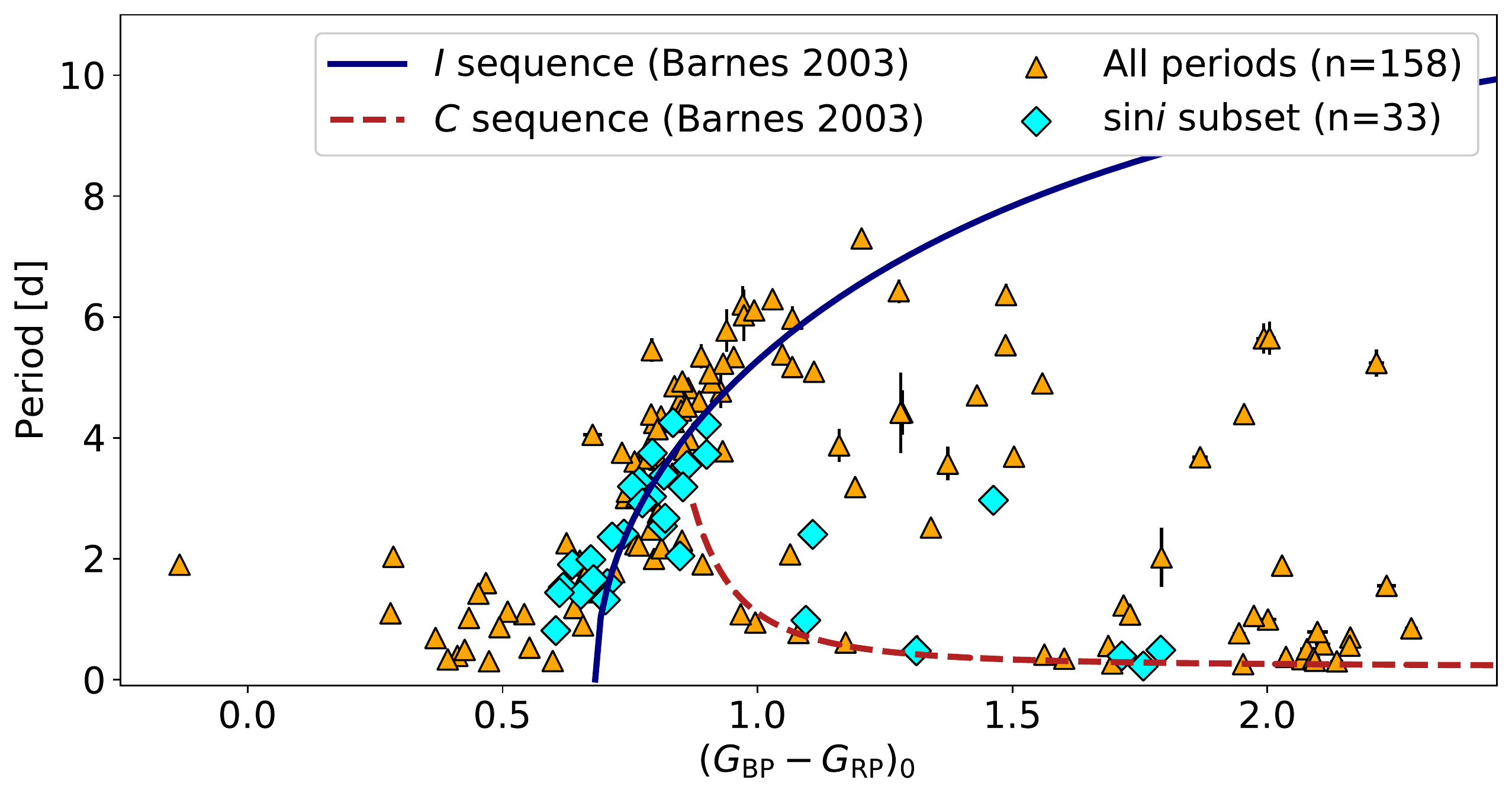}
    \caption{Color--period diagram of 158 measured NGC 2516 rotation periods plotted against dereddened $G_{\rm BP} - G_{\rm RP}$ color. Cyan points highlight the 33 stars for which we report $\sin i$ measurements. The $I$ sequence of slow rotators is traced by a 110 Myr isochrone calculated using Eqs.\ 1 and 2 of \citet[][]{barnes2003}. The bluest stars in the cluster are evolving off of the main sequence, and their periods are not predicted by this isochrone. The fast-rotating $C$ sequence of redder stars, observed in young clusters such as this one, is traced by Eq.\ 15 of \citet[][]{barnes2003}. Some stars in transition from the $C$ sequence to the $I$ sequence exist in the ``gap'' between the two.}
    \label{fig:rotperiods}
\end{figure*}

Fig.\ \ref{fig:cmd_results} shows the NGC 2516 color-magnitude diagram with these stars highlighted. Fig.\ \ref{fig:rotperiods} plots all measured periods as a function of dereddened $G_{\rm BP} - G_{\rm RP}$ color (determined using the extinction estimate from \citealt[][]{jackson2016}; see Sec.\ \ref{subsec:rvs}). The figure includes a 110 Myr color--period isochrone highlighting the $I$ sequence of slow rotators and another isochrone of the same age tracing the $C$ sequence of redder, rapid rotators \citep{barnes2003}. 

The mean period is 2.74 days, and the sample ranges from 0.22 to 7.30 days. The mean fractional uncertainty of the period measurements is $3.0 \%$. We detected rotation in 68 out of 238 stars with spectroscopic measurements (see Sec.\ \ref{subsec:rvs}). In the smaller sample of 93 stars with $v\sin i > 5$ km s$^{-1}$, we observed a rotation signature in 47 light curves. Among the stars for which we report $\sin i$, 31 out of 33 use period measurements consistent across all three of the \tess\ light-curve sets. The remaining two periods were detected by both PATHOS and CDIPS.

\begin{table*}
\begin{center}
\caption{Rotation periods of stars with $>$ 68\% confidence in NGC 2516 membership.}
\label{tab:periods}
\begin{tabular}{cccccc}
\hline
\hline
Gaia source ID & R.A.\ (deg) & Dec.\ (deg) & $P_{\rm mem}$ & $G_{\rm BP}-G_{\rm RP}$ & Period (days) \\
\hline
5290024533163062144 & 120.92 & -60.91 & 0.7 & -0.0077 $\pm$\ 0.0019 & 1.9006 $\pm$\ 0.0083 \\
5290715370058746368 & 118.83 & -60.96 & 0.8 & 0.4061 $\pm$\ 0.0018 & 1.096 $\pm$\ 0.017 \\
5289930181318610432 & 121.14 & -61.34 & 0.8 & 0.4116 $\pm$\ 0.0013 & 2.033 $\pm$\ 0.039 \\
5290671286515023744 & 119.38 & -60.99 & 0.9 & 0.4944 $\pm$\ 0.0014 & 0.69 $\pm$\ 0.14 \\
5290868820655092992 & 120.21 & -60.15 & 0.9 & 0.5183 $\pm$\ 0.0014 & 0.333 $\pm$\ 0.063 \\
5290739147002207232 & 119.26 & -60.61 & 0.8 & 0.5372 $\pm$\ 0.0022 & 0.3911 $\pm$\ 0.0083 \\
5290771204639131648 & 120.23 & -60.75 & 1.0 & 0.5516 $\pm$\ 0.0020 & 0.490 $\pm$\ 0.010 \\
5290738356723303936 & 119.35 & -60.65 & 0.9 & 0.5599 $\pm$\ 0.0067 & 1.019 $\pm$\ 0.029 \\
5291032132489758208 & 119.25 & -60.2 & 0.9 & 0.5785 $\pm$\ 0.0015 & 1.421 $\pm$\ 0.052 \\
5289934751163440000 & 121.16 & -61.25 & 0.8 & 0.5932 $\pm$\ 0.0014 & 1.596 $\pm$\ 0.020 \\
... & ... & ... & ... & ... & ... \\
\hline
\end{tabular}
\tablecomments{Table \ref{tab:periods} is published in its entirety in the machine-readable format.}
\end{center}
\end{table*}

\subsection{Inclination distribution}
\label{subsec:distribresults}
For 33 single-star members of NGC 2516, we report the most probable value for $\sin i$, along with 16th and 84th percentile error bars, in Table \ref{tab:inclinations}. The majority of these 33 stars are Sun-like, with the 26 bluest stars falling on the $I$ sequence of rotators. The 5 reddest stars populate the $C$ sequence, and the remaining 2 stars are in the ``gap'' between sequences (see Figs.\ \ref{fig:cmd_results} and \ref{fig:rotperiods}). Our $\sin i$ measurements have a mean fractional uncertainty of $\sim 8 \%$. When we break down this uncertainty by parameter, we find that $v \sin i$ makes the highest contribution with a mean fractional uncertainty of $\sim 5\%$. The mean uncertainties in $P$ and $R$ for these 33 stars are each $\sim 2\%$. 

\begin{figure*}
    \centering
    \includegraphics[scale=0.5]{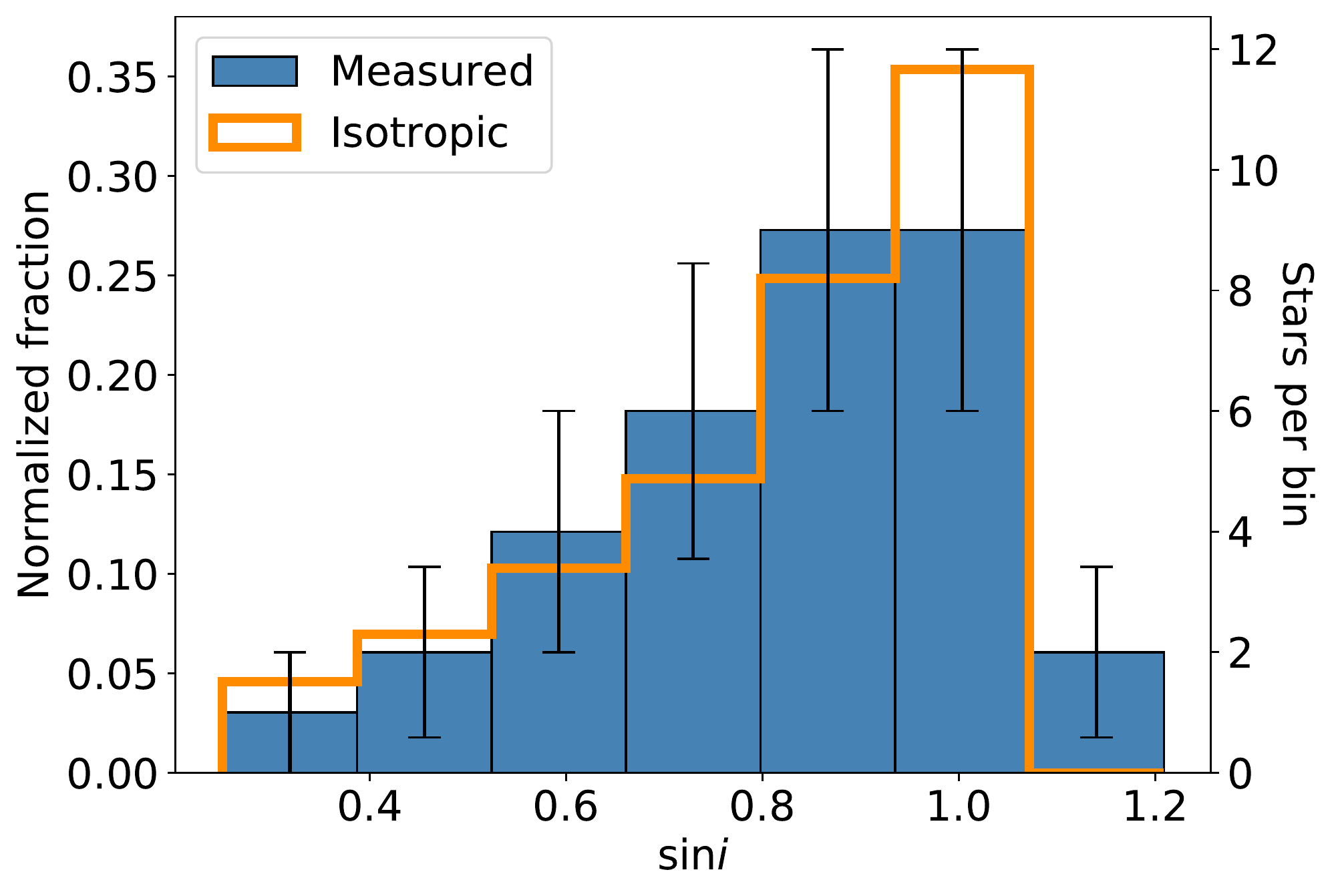}
    \caption{Histogram of 33 measured inclinations, with an isotropic distribution outlined in orange. The isotropic distribution does not contain values of $\sin i > 1$, while the maximum calculated $\sin i$ value is 1.134. Error bars assume a Poisson distribution of counts.}
    \label{fig:inclination_hist}
\end{figure*}

\begin{figure*}
    \centering
    \includegraphics[scale=0.5]{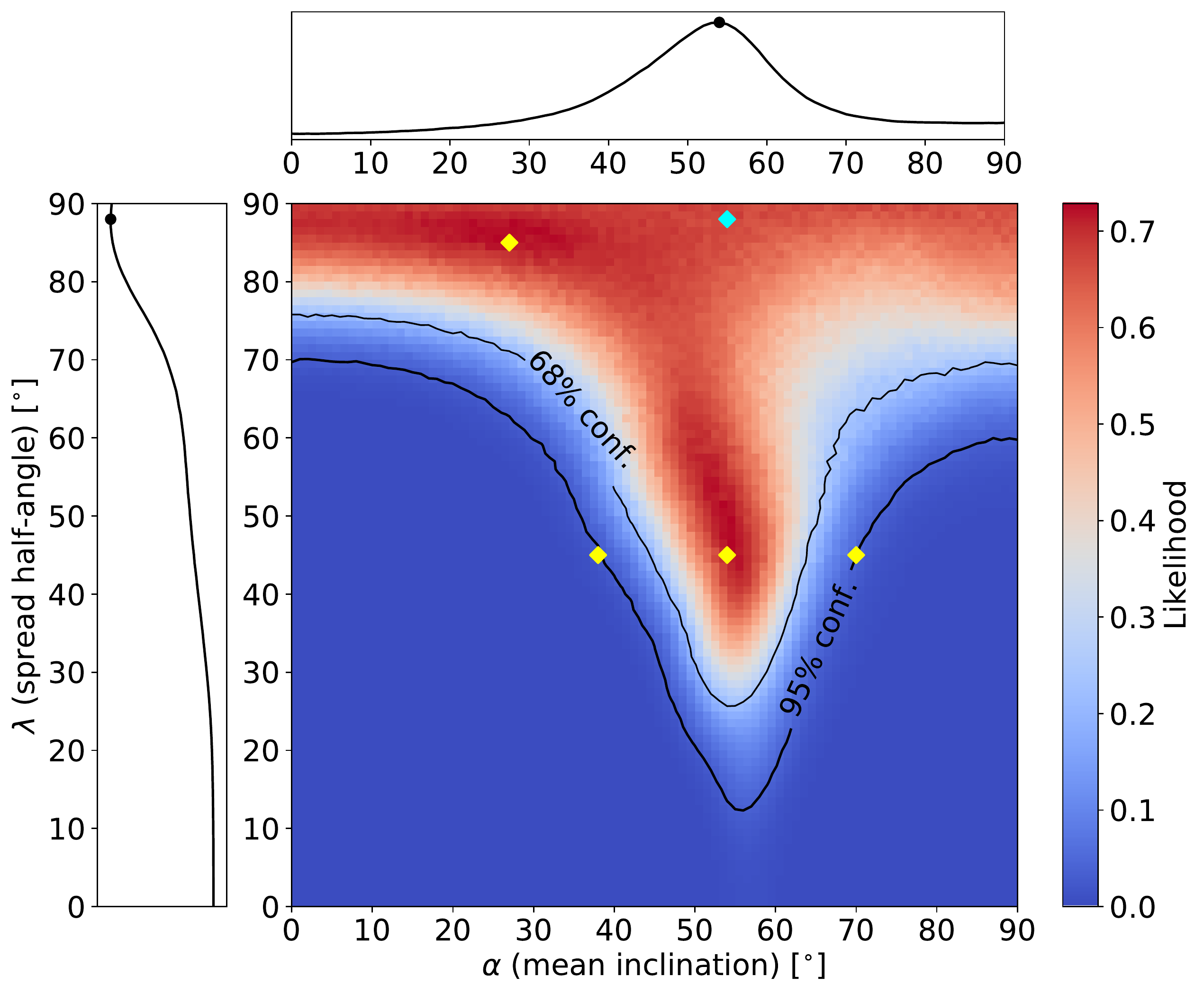}
    \caption{Visualization of cone model likelihood given $\alpha$ and $\lambda$, with marginalized PPDs for both parameters. High-spread half-angles are most probable, with a degenerate case of lower spreads possible for moderate mean inclinations. The cyan point highlights the most probable values of $\alpha$ and $\lambda$ based on their marginalized PPDs. Yellow points denote other combinations whose distributions are within the 95\% confidence interval for NGC 2516, including two degenerate peaks in the likelihood. Figs.\ \ref{fig:ecdffig} and \ref{fig:ecdffig_multi} show the cumulative distributions associated with the cyan and yellow points, respectively.}
    \label{fig:chi2fig}
\end{figure*}

\begin{figure}
    \centering
    \includegraphics[scale=0.45]{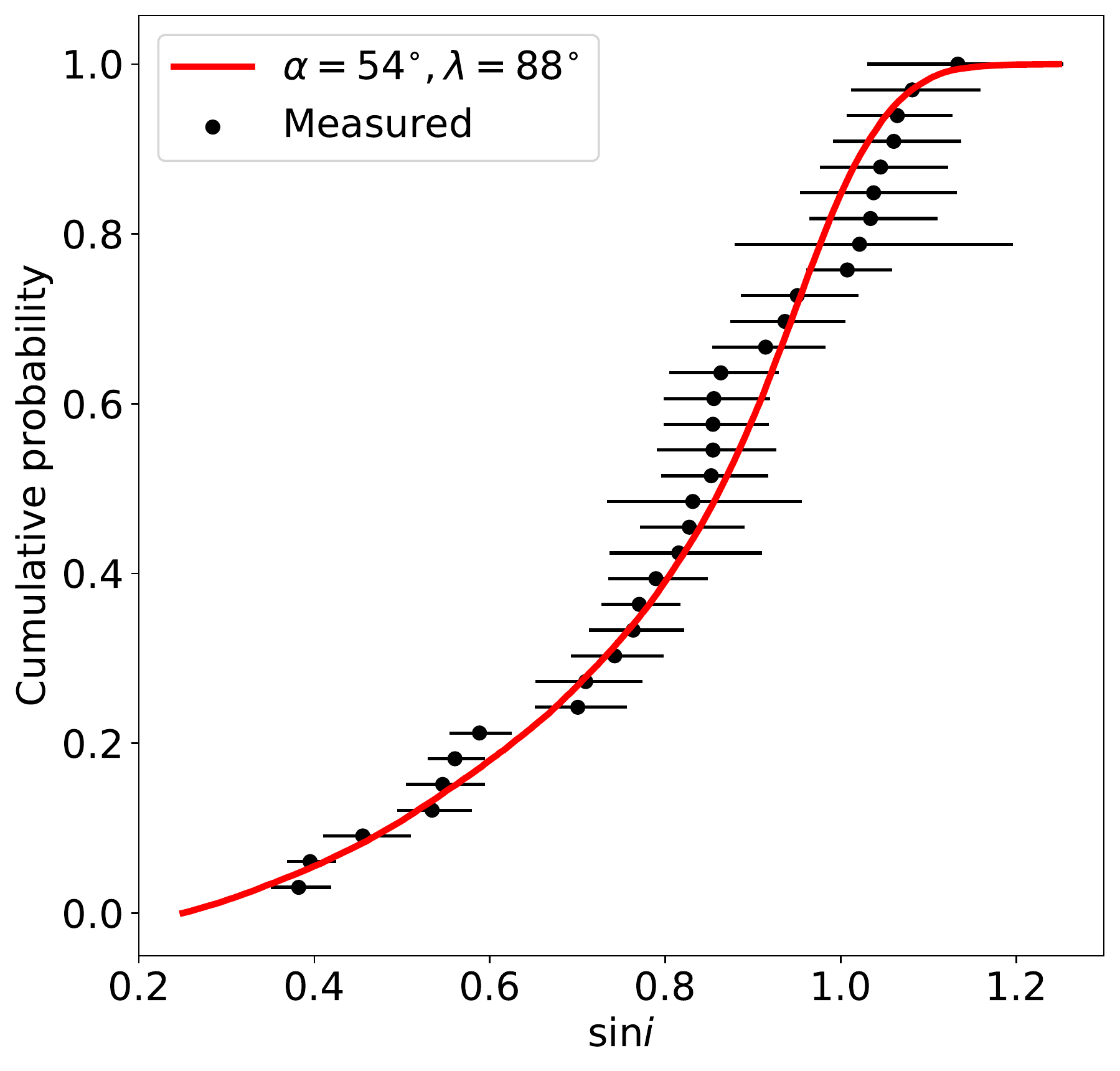}
    \caption{Empirical cumulative distribution of this work's $\sin i$ measurements and best-fit cone model distribution, corresponding to $\alpha = 54^{\circ}$ and $\lambda = 88^{\circ}$. The minimum detectable $\sin i$ is determined to be 0.25 (see Sec.\ \ref{subsec:distribution}). Measurements of $\sin i > 1$ are permitted, and we modeled them given the typical uncertainties of $v\sin i$, $P$ and $R$.}
    \label{fig:ecdffig}
\end{figure}

\begin{figure}
    \centering
    \includegraphics[scale=0.45]{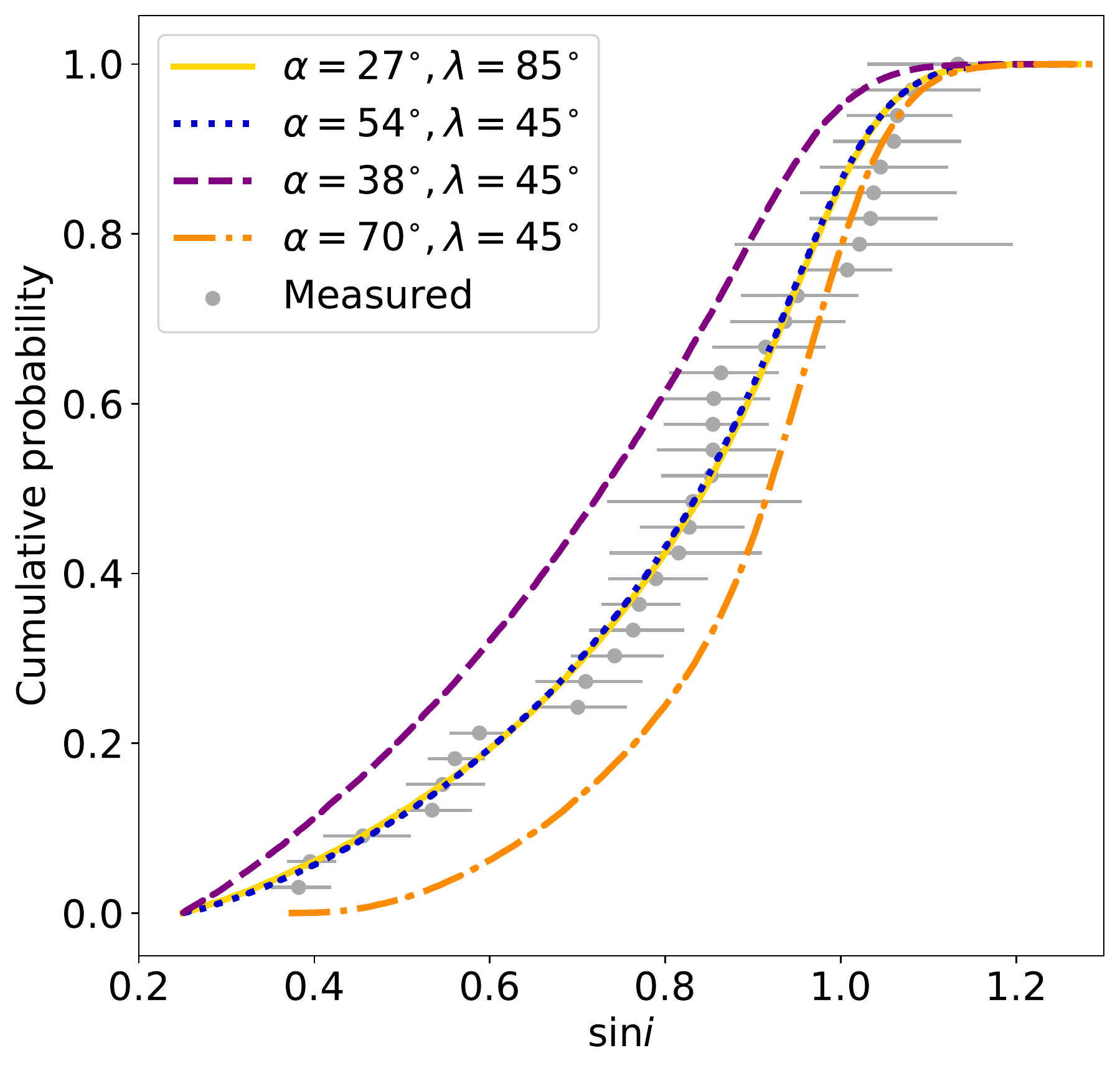}
    \caption{Empirical cumulative distribution of this work’s $\sin i$ measurements and selected distributions within the 95\% confidence interval for the cluster. The combinations of $\alpha$ and $\lambda$ shown here are marked by yellow diamonds in Fig.\ \ref{fig:chi2fig}. The degeneracy between isotropy and moderate alignment at moderate-spread half-angles is apparent in the overlap of the solid yellow and dotted blue curves.}
    \label{fig:ecdffig_multi}
\end{figure}

In Fig.\ \ref{fig:inclination_hist}  we compare the distribution of inclination measurements for NGC 2516 to a theoretical isotropic distribution, i.e.\ a uniform distribution in $\cos i$. To further examine our results, we show the likelihood of all combinations of the cone model's $\alpha$ and $\lambda$, with contours containing the 68\% and 95\% confidence intervals, in Fig.\ \ref{fig:chi2fig}. This figure also displays the marginalized PPDs for the two angles. The best fits favor high values of $\lambda$: $68 \%$ of the spread half-angle distribution is greater than $65^{\circ}$, and $95 \%$ is greater than $> 37^{\circ}$. The fits prefer intermediate values of $\alpha$, with $68 \%$ of the distribution falling between $32^{\circ}$ and $78^{\circ}$.
Based on the marginalized PPDs for each parameter, we found the best-fitting cone model to have an optimal $\sin i$ threshold of 0.25, mean inclination $\alpha = 54^{\circ}$, and spread half-angle $\lambda = 88^{\circ}$. We plot the empirical cumulative distribution of the data and the best-fit model in Fig.\ \ref{fig:ecdffig}. Fig.\ \ref{fig:ecdffig_multi} shows additional cumulative distribution functions for selected combinations of $\alpha$ and $\lambda$ within 95\% of the maximum likelihood, illustrating the degeneracy between an isotropic distribution and a moderately aligned one. We discuss this degeneracy further in Sec.\ \ref{subsec:discussion_isotropy}. Finally, we calculated an alignment coefficient (cf.\ Eq.\ \ref{eq:aligncoeff}) of $A = 0.29$ from our measurements.

\begin{table*}
\begin{center}
\caption{Inclinations and intermediate measurements for 33 NGC 2516 stars.}
\label{tab:inclinations}
\begin{tabular}{ccccccccc}
\hline
\hline
Gaia source ID & R.A.\ (deg) & Decl.\ (deg) & $P_{\rm mem}$ & $T_{\rm eff}$ (K) & Radius (\rsun) & $v\sin i$ (km s$^{-1}$) & Period (days) & $\sin i$ \\
\hline
5290715919819356800 & 118.96 & -60.97 & 0.7 & 4290 $\pm$\ 210 & 0.637 $\pm$\ 0.029 & 55.2 $\pm$\ 2.2 & 0.221 $\pm$\ 0.010 & 0.382$^{+0.037}_{-0.032}$ \\
5290830097230075904 & 119.97 & -60.37 & 0.9 & 6340 $\pm$\ 160 & 1.216 $\pm$\ 0.024 & 12.80 $\pm$\ 0.64 & 1.902 $\pm$\ 0.016 & 0.395$^{+0.030}_{-0.026}$ \\
5290667472588665600 & 119.65 & -61.08 & 0.7 & 5780 $\pm$\ 160 & 0.948 $\pm$\ 0.025 & 5.10 $\pm$\ 0.41 & 4.251 $\pm$\ 0.031 & 0.455$^{+0.055}_{-0.045}$ \\
5290723032285137024 & 119.25 & -60.83 & 0.8 & 4360 $\pm$\ 120 & 0.648 $\pm$\ 0.020 & 44.7 $\pm$\ 2.2 & 0.3905 $\pm$\ 0.0083 & 0.534$^{+0.045}_{-0.040}$ \\
5290754643245585792 & 120.3 & -60.9 & 0.9 & 5980 $\pm$\ 140 & 1.061 $\pm$\ 0.022 & 9.60 $\pm$\ 0.58 & 3.050 $\pm$\ 0.010 & 0.546$^{+0.048}_{-0.042}$ \\
5290737669533445248 & 119.14 & -60.65 & 0.7 & 6020 $\pm$\ 150 & 0.992 $\pm$\ 0.021 & 8.60 $\pm$\ 0.34 & 3.2738 $\pm$\ 0.0083 & 0.560$^{+0.034}_{-0.031}$ \\
5290667541308087168 & 119.65 & -61.05 & 0.7 & 5890 $\pm$\ 130 & 1.005 $\pm$\ 0.019 & 9.90 $\pm$\ 0.40 & 3.026 $\pm$\ 0.048 & 0.588$^{+0.037}_{-0.034}$ \\
5290817281048004736 & 119.81 & -60.67 & 0.9 & 5590 $\pm$\ 140 & 0.917 $\pm$\ 0.022 & 7.70 $\pm$\ 0.39 & 4.218 $\pm$\ 0.081 & 0.700$^{+0.056}_{-0.049}$ \\
5290717672165866496 & 119.05 & -60.85 & 0.9 & 4500 $\pm$\ 99 & 0.704 $\pm$\ 0.017 & 8.50 $\pm$\ 0.51 & 2.965 $\pm$\ 0.046 & 0.709$^{+0.065}_{-0.057}$ \\
5290720042987976576 & 119.3 & -60.85 & 0.8 & 6070 $\pm$\ 160 & 1.067 $\pm$\ 0.022 & 10.70 $\pm$\ 0.54 & 3.746 $\pm$\ 0.048 & 0.742$^{+0.056}_{-0.050}$ \\
5290747977456314880 & 120.12 & -61.01 & 0.7 & 5980 $\pm$\ 170 & 1.033 $\pm$\ 0.024 & 12.50 $\pm$\ 0.63 & 3.1928 $\pm$\ 0.0083 & 0.763$^{+0.058}_{-0.050}$ \\
5290667713106775040 & 119.6 & -61.06 & 0.9 & 5910 $\pm$\ 120 & 1.022 $\pm$\ 0.019 & 11.80 $\pm$\ 0.47 & 3.381 $\pm$\ 0.044 & 0.770$^{+0.047}_{-0.043}$ \\
5290713652076611200 & 118.77 & -61.07 & 0.9 & 5800 $\pm$\ 140 & 0.999 $\pm$\ 0.024 & 10.70 $\pm$\ 0.54 & 3.726 $\pm$\ 0.031 & 0.789$^{+0.059}_{-0.054}$ \\
5290710800218328192 & 118.83 & -61.07 & 0.7 & 5870 $\pm$\ 190 & 1.054 $\pm$\ 0.021 & 12.20 $\pm$\ 0.98 & 3.542 $\pm$\ 0.023 & 0.815$^{+0.095}_{-0.079}$ \\
5290728834785867264 & 118.71 & -60.84 & 0.8 & 5980 $\pm$\ 150 & 1.047 $\pm$\ 0.022 & 15.00 $\pm$\ 0.75 & 2.921 $\pm$\ 0.048 & 0.827$^{+0.063}_{-0.056}$ \\
5290725231308404864 & 119.43 & -60.74 & 0.9 & 6200 $\pm$\ 140 & 1.124 $\pm$\ 0.026 & 19.20 $\pm$\ 0.96 & 2.41 $\pm$\ 0.24 & 0.831$^{+0.12}_{-0.098}$ \\
5290739937276199168 & 119.3 & -60.55 & 0.7 & 5070 $\pm$\ 110 & 0.799 $\pm$\ 0.019 & 35.2 $\pm$\ 1.8 & 0.9792 $\pm$\ 0.0083 & 0.852$^{+0.065}_{-0.057}$ \\
5290723204083832448 & 119.29 & -60.8 & 0.9 & 6320 $\pm$\ 140 & 1.207 $\pm$\ 0.026 & 32.8 $\pm$\ 1.6 & 1.587 $\pm$\ 0.050 & 0.854$^{+0.072}_{-0.064}$ \\
5290725781064133760 & 119.29 & -60.74 & 0.7 & 5840 $\pm$\ 140 & 0.966 $\pm$\ 0.022 & 13.10 $\pm$\ 0.66 & 3.188 $\pm$\ 0.013 & 0.854$^{+0.064}_{-0.056}$ \\
5290770345645486976 & 119.97 & -60.69 & 0.9 & 6300 $\pm$\ 130 & 1.193 $\pm$\ 0.024 & 21.9 $\pm$\ 1.1 & 2.358 $\pm$\ 0.028 & 0.855$^{+0.064}_{-0.057}$ \\
5290838962037067648 & 119.55 & -60.42 & 1.0 & 6430 $\pm$\ 140 & 1.26 $\pm$\ 0.024 & 68.0 $\pm$\ 3.4 & 0.809 $\pm$\ 0.016 & 0.863$^{+0.066}_{-0.059}$ \\
5290771032840440832 & 120.26 & -60.75 & 0.8 & 6410 $\pm$\ 140 & 1.257 $\pm$\ 0.026 & 38.1 $\pm$\ 1.9 & 1.526 $\pm$\ 0.016 & 0.914$^{+0.068}_{-0.061}$ \\
5290777522534884864 & 120.35 & -60.59 & 0.9 & 6460 $\pm$\ 140 & 1.281 $\pm$\ 0.025 & 43.1 $\pm$\ 2.2 & 1.408 $\pm$\ 0.010 & 0.936$^{+0.069}_{-0.062}$ \\
5291030448862535808 & 119.24 & -60.3 & 0.7 & 6230 $\pm$\ 160 & 1.152 $\pm$\ 0.025 & 27.9 $\pm$\ 1.4 & 1.984 $\pm$\ 0.016 & 0.950$^{+0.070}_{-0.064}$ \\
5290673348103622272 & 119.49 & -60.89 & 0.9 & 4960 $\pm$\ 140 & 0.808 $\pm$\ 0.020 & 17.20 $\pm$\ 0.52 & 2.4006 $\pm$\ 0.0083 & 1.008$^{+0.051}_{-0.047}$ \\
5290653075858442752 & 119.63 & -61.17 & 0.8 & 4520 $\pm$\ 150 & 0.703 $\pm$\ 0.019 & 75.0 $\pm$\ 9.0 & 0.4744 $\pm$\ 0.0083 & 1.022$^{+0.18}_{-0.14}$ \\
5290824320493640576 & 120.14 & -60.52 & 0.8 & 6260 $\pm$\ 160 & 1.181 $\pm$\ 0.027 & 46.9 $\pm$\ 2.3 & 1.3173 $\pm$\ 0.0083 & 1.034$^{+0.077}_{-0.070}$ \\
5290664929967787264 & 119.26 & -61.03 & 1.0 & 6490 $\pm$\ 130 & 1.375 $\pm$\ 0.030 & 50.0 $\pm$\ 2.5 & 1.438 $\pm$\ 0.063 & 1.038$^{+0.095}_{-0.084}$ \\
5290826936134381440 & 120.13 & -60.46 & 0.8 & 6280 $\pm$\ 150 & 1.175 $\pm$\ 0.026 & 37.6 $\pm$\ 1.9 & 1.654 $\pm$\ 0.010 & 1.046$^{+0.077}_{-0.069}$ \\
5290652938419483904 & 119.59 & -61.2 & 0.9 & 6010 $\pm$\ 170 & 1.031 $\pm$\ 0.020 & 21.8 $\pm$\ 1.1 & 2.5387 $\pm$\ 0.0083 & 1.061$^{+0.077}_{-0.069}$ \\
5290814807146918016 & 119.65 & -60.78 & 0.9 & 5720 $\pm$\ 130 & 0.985 $\pm$\ 0.024 & 19.90 $\pm$\ 0.60 & 2.668 $\pm$\ 0.074 & 1.065$^{+0.063}_{-0.058}$ \\
5290715954179096320 & 119.01 & -60.96 & 0.7 & 5950 $\pm$\ 140 & 1.068 $\pm$\ 0.020 & 28.6 $\pm$\ 1.4 & 2.0446 $\pm$\ 0.0083 & 1.082$^{+0.078}_{-0.070}$ \\
5290744983857413376 & 120.12 & -61.13 & 0.8 & 4080 $\pm$\ 170 & 0.583 $\pm$\ 0.022 & 68.4 $\pm$\ 4.1 & 0.486 $\pm$\ 0.017 & 1.134$^{+0.12}_{-0.10}$ \\
\hline
\end{tabular}
\tablecomments{Table \ref{tab:inclinations} is also available in the machine-readable format.}
\end{center}
\end{table*}

\begin{table*}
\begin{center}
\caption{Mean measurements of NGC 2516 internal kinematics.}
\label{tab:rotation}
\begin{tabular}{ccccc}
\hline
\hline
 & $\theta_{\rm c}$ (deg) & $v_{0}$ (km s$^{-1}$) & $v_{\rm rot}$ (km s$^{-1}$) & $\sigma$ (km s$^{-1}$) \\
\hline
Tangential & --- & $0.035 \pm 0.033$ & --- & $0.752 \pm 0.024$ \\
Radial & --- & $-0.331 \pm 0.032$ & --- & $ 0.718 \pm 0.023$ \\
LOS & $2 \pm 70$ & $23.799 \pm 0.066$ & 0.083$_{-.057}^{+.083}$ & 
$0.939 \pm 0.051$ \\
\hline
\end{tabular}
\end{center}
\end{table*}

\subsection{Cluster rotation}
In the LOS direction, we measured the cluster's rotation speed to be $0.083_{-0.057}^{+0.083}$ km s$^{-1}$, greater than zero at the 1.5$\sigma$ level. We constrained the position angle of the rotation axis to be $2^{\circ} \pm 70^{\circ}$. We refined the cluster's recessional velocity to a value of $23.799 \pm 0.066$ km s$^{-1}$, and we measured the LOS velocity dispersion to be $0.939 \pm 0.051$ km s$^{-1}$. Table \ref{tab:rotation} presents all of our mean kinematic measurements for NGC 2516.

\begin{figure}
    \centering
    \includegraphics[scale=0.48]{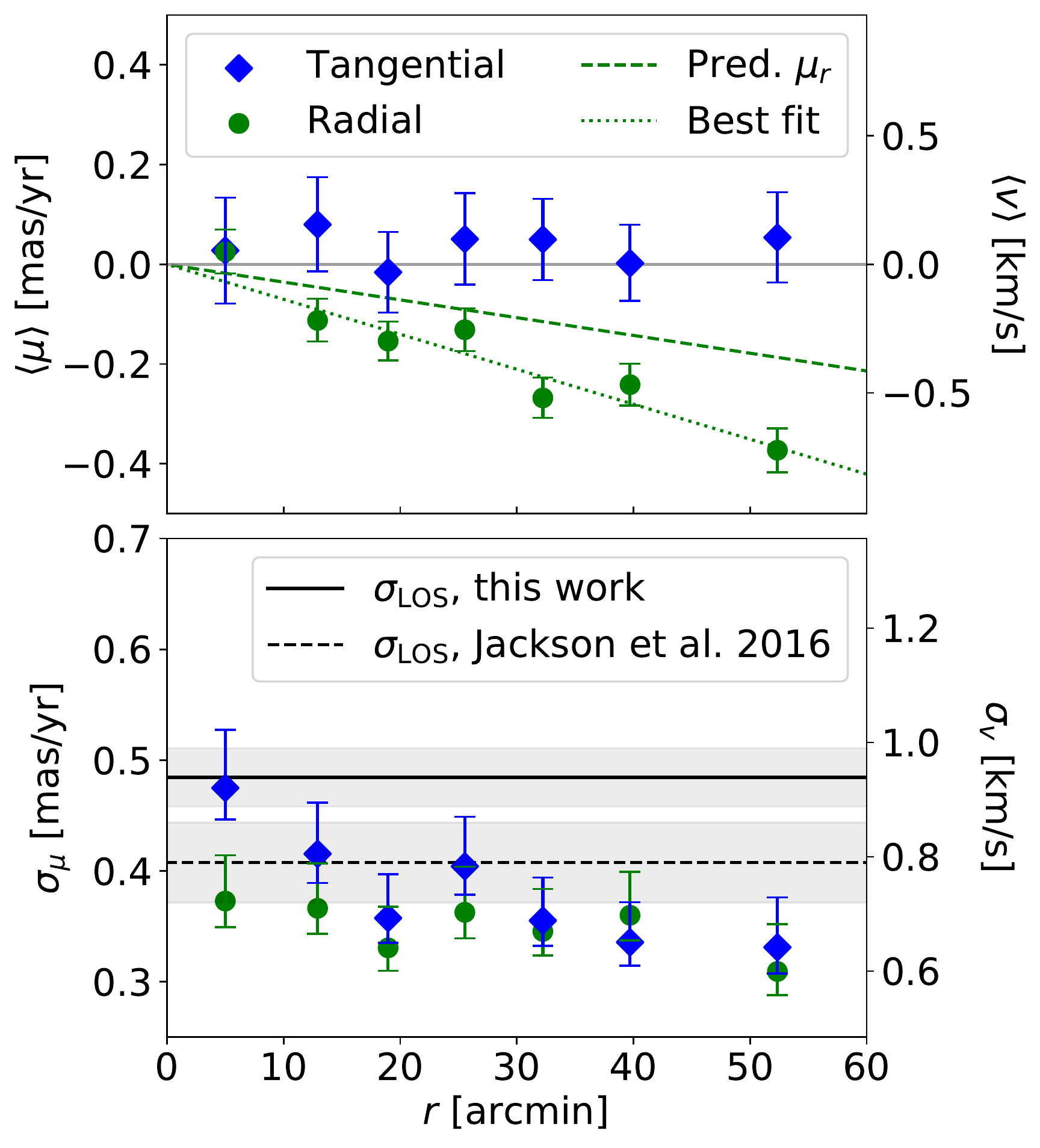}
    \caption{{Top:} mean tangential (blue diamond) and radial (green circle) proper motions of NGC 2516 stars versus radial distance $r$ from cluster center. The predicted apparent radial contraction due to the cluster's recession speed is plotted as the dashed line. The best linear fit to the radial points passing through (0,0) is traced by the dotted line. The observed trend is discrepant from the prediction by a factor of $\sim$ 2. {Bottom:} velocity dispersions in the tangential and radial directions, compared with the cluster's LOS velocity dispersion (black) as measured by this work and \citet[][]{jackson2016}. Shaded regions indicate 68\% confidence intervals.}
    \label{fig:cluster_rot}
\end{figure}

We plot the radially binned proper-motion results in Fig.\ \ref{fig:cluster_rot}. The top panel of the figure shows the radial dependence of $\mu_{\theta}$ and $\mu_{r}$, the tangential and radial components of proper motion, respectively. The cluster stars' motion in the tangential direction is only 1.1$\sigma$ greater than zero, and we found no significant trend in $v_{\theta}$ with distance from the center. We found a trend of decreasing $\mu_{r}$ with increasing radial distance from cluster center. For comparison, we plot both the best linear fit to the data points and the predicted radial dependence of the apparent contraction caused by the cluster's motion away from Earth (calculated from Eq.\ 6 of \citealt[][]{vandeven2006}). The best fit and the predicted radial motion are discrepant, with the stars' observed radial motion larger than the prediction by a factor of $\sim 2$.

The bottom panel of Fig.\ \ref{fig:cluster_rot} shows the proper-motion-derived velocity dispersion in the tangential and radial directions. For comparison with the plane-of-sky directions, we plot two values for the cluster's LOS velocity dispersion: this work's measurement and one from \citet[][]{jackson2016} using the same dataset. We discuss their discrepancy in Sec.\ \ref{subsec:discussion_kinematics}. Stars closest to the center of the cluster show a velocity anisotropy such that the dispersion in the radial direction is less than those of the tangential and LOS directions. At greater distances from the cluster center, the tangential and radial dispersions are in statistical agreement with each other and the LOS dispersion from \citet[][]{jackson2016}, but they show discrepancy with our LOS dispersion measurement.

\section{Discussion} \label{sec:discussion}
\subsection{Rotation periods}
\label{subsec:discussion_periods}
Our rotation period measurements agree with the color--period predictions for the populations of slow (P $\gtrsim$ 1 day), Sun-like $I$ sequence rotators and fast (P $\lesssim$ 1 day), redder stars on the $C$ sequence. The empirical color--period isochrones of \citet[][]{barnes2003} are a good fit to the cluster's $I$ and $C$ sequences at an age of 110 Myr, similar to most previous age estimates. One exception is the age of $\sim 250$ Myr reported by \citet[][]{bossini2019}. Using this age to calculate the color--period isochrones results in slower-rotating sequences that do not agree with the rotation rates determined here. The roughly few percent differences in periods measured using multiple data sets are consistent with the typical reported period uncertainty of $\sim 3\%$.

If starspots exist across a range of stellar latitudes, as they do on the Sun, then ignoring differential rotation may systematically overestimate photometric rotation periods.
We approximate the potential contribution of differential rotation to our period measurements using the model of \citet[][]{reiners2003_diffrot}. This model, based on observations of the Sun, describes the dependence of a star's angular rotation rate $\Omega$ on the latitude $l$ with the equation
\begin{equation}
\label{eq:diffrot}
    \Omega(l) = \Omega_{\rm{eq}} (1 - \alpha \sin^2 l),
\end{equation}
where $\Omega_{\rm{eq}}$ is the equatorial angular rotation rate and $\alpha$ is the fractional difference in rotation rate between the equator and the poles. For Sun-like differential rotation, $\alpha \sim 0.2$ \citep{reiners2003_diffrot} and spots exist at latitudes where $ 5^{\circ} \lesssim |l| \lesssim 40^{\circ}$ \citep[e.g.\ Fig.\ 1 of][]{hathaway2011_sunspotlatitudes}.

Assuming Sun-like differential rotation for our targets, we examine a worst-case scenario for systematic error if spots are located only at $|l| \sim 40^{\circ}$. The sinusoidal modulation caused by these spots would yield a measured rotation period $\sim 8\%$ greater than the equatorial period according to Eq. \ref{eq:diffrot}.
A mitigating factor in our inclination analysis is that the GES and GALAH $v\sin i$ measurements for NGC 2516 do not account for differential rotation and thus may be underestimated on the order of $10\%$ \citep{hirano2012}. Inclinations determined by Eq.\ \ref{eq:spectrophot} are proportional to the product of $v\sin i$ and $P$, so the systematics should partially cancel out.

\subsection{Stellar spins}
\label{subsec:discussion_isotropy}
With a 95\% confidence spread half-angle constraint of $\lambda > 37^{\circ}$ from Sec.\ \ref{subsec:distribresults}, we find evidence for either isotropic or moderately aligned stellar spins in NGC 2516. The alignment coefficient $A = 0.29$ is consistent with isotropy but does not exclude all possible anisotropic scenarios. The measured $\sin i$ distribution in Fig.\ \ref{fig:inclination_hist} is consistent with the distribution of isotropic spins. However, Fig.\ \ref{fig:chi2fig} shows a ``peninsula'' of high likelihood at moderate ranges of $\alpha$ and $\lambda$. The presence of this feature illustrates the possibility of a tighter alignment of spins for intermediate mean inclinations. As shown by Fig.\ 1 of \citet{jackson2010}, even a perfectly isotropic distribution of spins is degenerate with moderate alignment.

Numerical simulations by \citet{corsaro2017} predicted that spin alignment can occur in stars with mass $> 0.7$ \msun. Most of the stars for which we measured $\sin i$ are Sun-like and therefore more massive than that threshold. An isotropic result for these stars would suggest that turbulence dominated the kinetic energy of the cluster's progenitor molecular cloud over ordered rotation. In light of these simulations, such a result would suggest that the energy from rotation $E_{\rm{rot}}$ was smaller than the energy $E_{\rm{tur}}$ in turbulent motions to the point of being negligible. An isotropic result would also imply that turbulence misaligns protostellar cores from their magnetic fields, allowing massive disks to form in the absence of strong magnetic braking. A moderately anisotropic result would suggest a greater contribution of $E_{\rm rot}$ to the energy balance, but $E_{\rm tur}$ could still be dominant ($\sim 10$ times greater) in this scenario. Additional characterizations of cluster spin distributions should facilitate more definitive physical interpretations, as a greater sample size may reveal patterns or outliers.

\subsection{Selection effects from $v \sin i$ threshold}
Our requirement of $v\sin i > 5$ km s$^{-1}$ for all inclination measurements excludes both low-inclination and slower-rotating stars from our analysis. To quantify these selection effects on our NGC 2516 sample, we performed Monte Carlo simulations of $\sin i$ measurements using our $v\sin i$ threshold and assuming a degenerate isotropic/moderately aligned inclination distribution function similar to that of NGC 2516.

We used the 33 NGC 2516 stars with period, effective temperature, radius, and $\sin i$ measurements to provide empirical relations applicable to a simulated population of stars. 
We performed smoothed polynomial spline fits relating our $T_{\rm eff}$ and $R$ measurements to $G_{\rm BP} - G_{\rm RP}$ color. We then selected color ranges for the $I$ and $C$ rotation sequences based on visual inspection of the color--period diagram for the 33 stars, interpolating the 110 Myr \citet[][]{barnes2003} period predictions across a grid of $10^6$ simulated stars with \gaia\ colors in the range of $0.698 \leq G_{\rm BP}-G_{\rm RP} \leq 1.803$. We sampled an equal number of inclinations from the assumed distribution function to calculate simulated $v\sin i$ measurements, and we then excluded all values less than 5 km s$^{-1}$. We also incorporated theoretical rotation tracks from \citet[][]{vsp2013_rotation_predictions} at 550 Myr, the age at which stars begin to be described by a single period--mass relation. We used the solar-metallicity, ``slow launch'' period predictions to provide both an upper limit and an extended look at selection biases due to a $v\sin i$ threshold. 

\begin{figure*}
    \centering
    \includegraphics[scale=0.45]{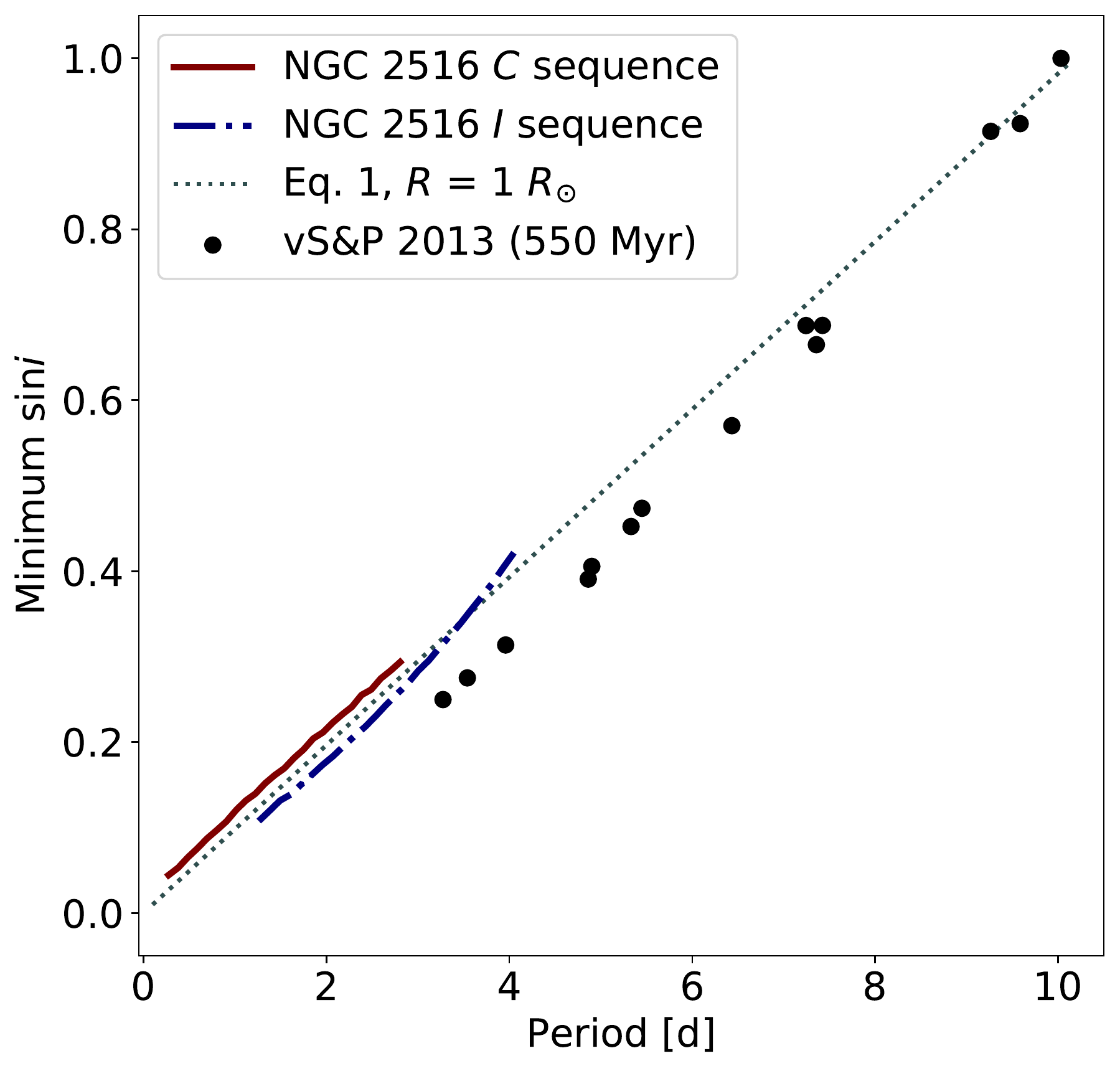}
    \includegraphics[scale=0.45]{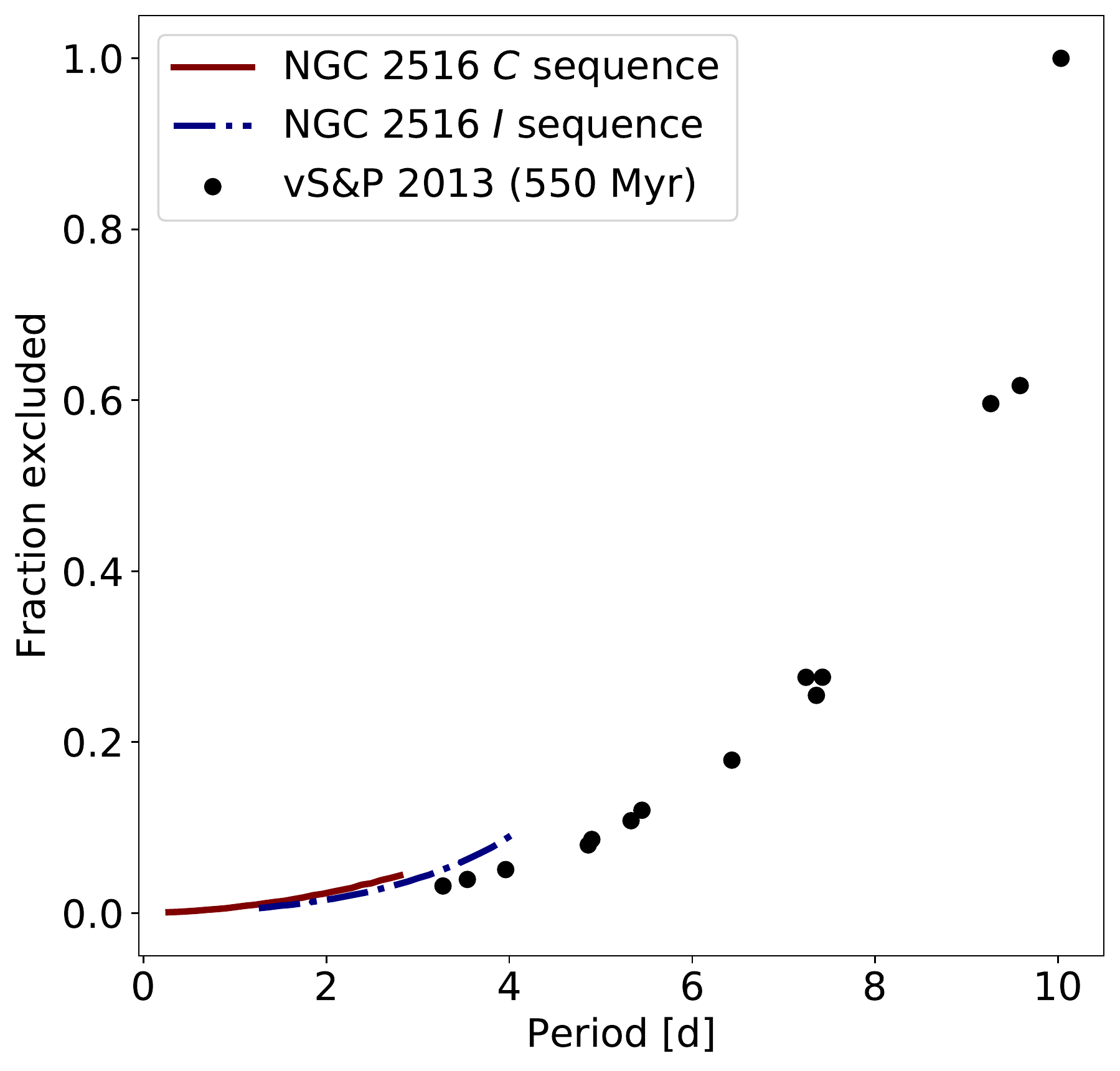}
    \caption{Monte Carlo simulation results showing the period-dependent effect of requiring $v\sin i > 5$ km s$^{-1}$. The NGC 2516 simulations use period predictions from 110 Myr \citet[][]{barnes2003} $I$ and $C$ sequence isochrones. We also incorporate a 550 Myr theoretical track from  \citet{vsp2013_rotation_predictions}, labeled ``vS\&P 2013'' above. {Left:} minimum measurable $\sin i$, with an analytic estimate using Eq.\ \ref{eq:spectrophot} evaluated at $v\sin i = 5$ km s$^{-1}$ and $R = 1$ \rsun. {\it Right:} fraction of stars excluded from further analysis owing to the threshold, assuming an isotropic/moderately aligned inclination distribution corresponding to NGC 2516. All vS\&P 2013 points with periods $>10$ days, including the upper rightmost point in this plot, are completely excluded by the $v\sin i$ threshold.}
    \label{fig:vsini_montecarlo_period}
\end{figure*}

\begin{figure*}
    \centering
    \includegraphics[scale=0.45]{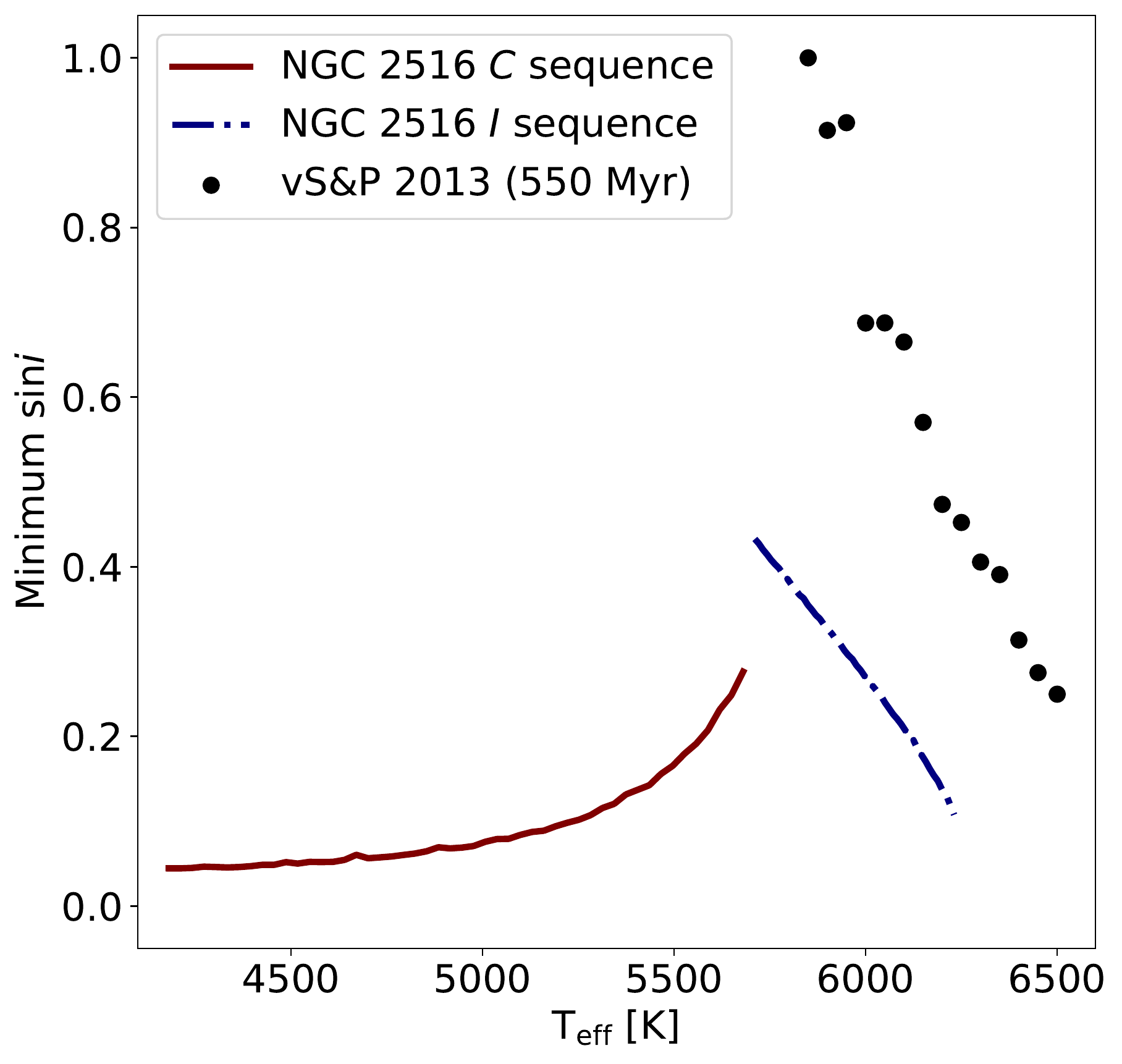}
    \includegraphics[scale=0.45]{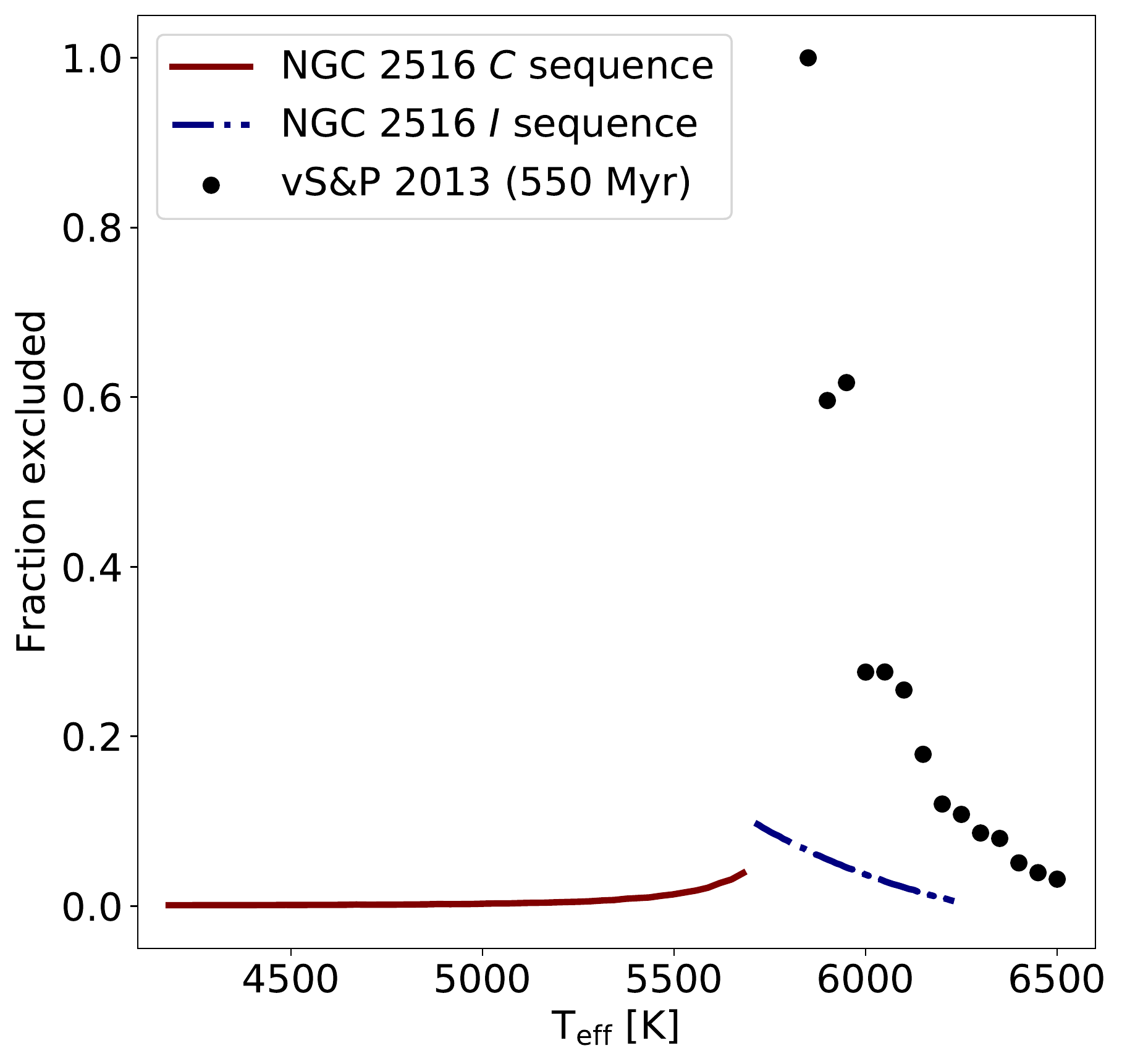}
    \caption{The same simulation conditions as Fig.\ \ref{fig:vsini_montecarlo_period}, but with results plotted against effective temperature. The most exclusion in the NGC 2516 sample due to the $v\sin i$ threshold occurs for $\sim 10\%$ of K-type stars, most of which rotate at the slowest rates of the $I$ sequence. The 550 Myr results predict more severe exclusion of stars in older clusters. All vS\&P 2013 points with $\textrm{T}_{\rm eff} < 5850$ K, including the highest point in this plot, are completely excluded by the $v\sin i$ threshold.}
    \label{fig:vsini_montecarlo_teff}
\end{figure*}

Figs.\ \ref{fig:vsini_montecarlo_period} and \ref{fig:vsini_montecarlo_teff} show the minimum observable $\sin i$ and fraction of stars excluded owing to a $v\sin i > 5$ km s$^{-1}$ threshold as a function of rotation period and effective temperature, respectively. With a typical minimum $\sin i$ of 0.17, very few simulated $C$ sequence stars (0.1 - 0.4\%) are excluded owing to their $v\sin i$ measurement. These stars' intrinsic faintness, however, makes it difficult to obtain high signal-to-noise ratio observations of their spectra, explaining the dominance of brighter $I$ sequence rotators in our 33-star sample. The simulated $I$ sequence has a typical $\sin i$ threshold of 0.25, in agreement with our cone model fit. Between 0.6\% and 10\% of these stars are excluded by the $v\sin i$ threshold, with K dwarfs experiencing the most prominent exclusion. Thus, the bias is strongest against stars near the minimum stellar mass (0.7 \msun) predicted by \citet[][]{reyraposo2018} to potentially display spin alignment. We know of no predictions of mass-dependent mechanisms of spin alignment besides this cutoff, so we do not expect the color-dependent consequences of the $v\sin i$ threshold to bias our determination of the spin distribution of the 33 stars selected.

The simulation results for the 550 Myr track from \citet[][]{vsp2013_rotation_predictions} show a greater level of exclusion. Stars with $P > 10$ days and $T_{\rm eff} < 5850$ K are completely excluded by a $v\sin i < 5$ km s$^{-1}$ threshold owing to their slow rotation. Only stars with $P \lesssim 6$ days and $T_{\rm eff} \gtrsim 6250$ K yield levels of exclusion similar to NGC 2516. The younger age of NGC 2516 by a factor of $\sim$ 5 spares most of its stars from these strict thresholds. However, these results highlight the limitations of using the spectrophotometric method to measure low inclinations in older clusters. Other techniques such as asteroseismology can complement our method by determining the inclinations of stars in such clusters without the selection bias against pole-on rotators.

\subsection{Cluster rotation}
\label{subsec:discussion_kinematics}
One must be mindful of systematic errors in Gaia DR2 astrometry when analyzing proper-motion data \citep[][]{gaiaclusters2018}. \citet{vasiliev2019} found that measurements of a cluster's tangential motion must be greater than a $\sim 0.05$ mas yr$^{-1}$ systematic floor to be considered significant. Our mean tangential velocity measurement of 0.035 $\pm$ 0.033 km s$^{-1}$ (0.018 $\pm$ 0.017 mas yr$^{-1}$) is therefore not large enough to stand above potential systematic errors. In addition, our analysis of the cluster's LOS velocities suggests overall rotation only at the 1.5$\sigma$ level. The plane-of-sky and LOS rotation measurements do not provide sufficient evidence for overall NGC 2516 rotation due to their small statistical significance.

The decreasing trend of radial proper motions in NGC 2516, indicative of apparent contraction, is qualitatively consistent with a cluster receding away from Earth, as NGC 2516 is doing. However, the significant quantitative discrepancy between predicted and observed values calls for an additional explanation. The discrepancy may be caused by \gaia\ systematics, encountered by \citet{kamann2019} in a similar analysis of cluster rotation. We calculated a total expected systematic error of $\sim 0.03$ mas yr$^{-1}$ using data from Sec.\ 5.4 of \citet{lindegren2018} and Eq.\ 4 of \citet[][]{kamann2019}, and we verified this quantity for a cluster of $\sim 1^{\circ}$ angular size in Fig.\ 3 of \citet[][]{vasiliev2019}. This level of systematic error still does not explain the radial proper-motion discrepancies at large distances ($>30$') from cluster center, which are 3-6 times larger than the predicted systematics.

It is possible that the excess apparent contraction is due to mass segregation within the cluster, as more massive stars move inward toward the center. In another study of cluster proper motions, \citet[][]{bonatto2011} suggested that higher-than-expected proper motions might be attributed to large-scale mass segregation. The relaxation timescale and age of NGC 2516 are both of order 100 Myr based on mass estimates for the cluster \citep{jeffries2001}, supporting this possibility.

Our measurement of the cluster's LOS velocity dispersion is mostly inconsistent with the dispersions we determined from proper motions in the tangential and radial directions. Our mean plane-of-sky dispersion measurements do agree, however, with the smaller LOS dispersion from \citet[][]{jackson2016}. The \citeauthor[][]{jackson2016} measurement is based on the same GES dataset, although \gaia\ astrometry was not yet available to determine cluster membership. It is likely that NGC 2516 contains many binary systems that introduce additional spread into the LOS velocities through their orbital motion. \citet[][]{sollima2010} estimated a $\sim 66 \%$ binary fraction in the cluster, with a minimum of $25 \%$. \citet[][]{bianchini2016} found that for globular clusters a binary fraction of $\sim 50 \%$ may induce a systematic bias in a cluster's velocity dispersion of order 0.1-0.3 km s$^{-1}$. In addition, \citet[][]{geller2015} calculated a dispersion correction of $\sim 0.2$ km s$^{-1}$ due to unresolved binaries in the open cluster M67, with an estimated binary fraction of 57\%.

The dispersion measurement of \citeauthor[][]{jackson2016} accounted for this bias through an estimate of the cluster's binary fraction and assumptions about the binary period and mass ratio distributions. To limit our dependence on such assumptions, we did not incorporate them into our analysis of the LOS velocities. The difference between the our LOS dispersion measurement and that of \citeauthor[][]{jackson2016} is of the same order of magnitude as the estimated contribution of binary velocities. Therefore, the discrepancy of our LOS velocity dispersion with the plane-of-sky dispersions should not be interpreted as indicative of significant anisotropic kinematics. 

\section{Conclusion} \label{sec:conclusion}
Motivated by predictions of numerical simulations and conflicting observational results on stellar spin axis distributions in open clusters, we performed a detailed study of stellar spin in NGC 2516. Starting with 535 likely cluster members on the single-star main sequence, we synthesized data from ground- and space-based telescopes to measure 158 rotation periods and 33 projected inclinations. 

Our rotation period measurements are fit well by gyrochronology predictions at an age of 110 Myr. We found that the cluster's inclination distribution favors isotropy or moderate alignment among the stars' spin axes, with a spread half-angle $\lambda > 65^{\circ}$ with 68\% confidence and $> 37^{\circ}$ at 95\% confidence. Our three-dimensional analysis of proper motions and LOS velocities did not provide support for overall cluster rotation. We detected a significant trend in the cluster's radial motion that cannot be geometrically explained by its recessional velocity along the LOS or \gaia\ systematic errors. We interpret the trend as evidence of ongoing mass segregation in NGC 2516.

Time-series photometry from \tess\ has enabled the study of open cluster rotation across most of the sky. Survey magnitudes paired with \gaia\ parallaxes facilitate the precise determination of stellar radii. Data from Gaia EDR3 will offer improved astrometric and photometric measurements that will refine cluster membership and distances. Subsequently, the full Gaia DR3 will provide new spectroscopic insight: Blue/Red Photometer spectra will help constrain the effective temperature of target stars, while Radial Velocity Spectrometer (RVS) spectra will identify binary stars and yield radial velocities for studies of cluster kinematics. While $v\sin i$ data from RVS will be limited in precision to $\gtrsim 10$ km s$^{-1}$ \citep{gomboc2005_gaiavsini}, the public release of the full Gaia-ESO survey will refine the $v\sin i$ measurements used in this study.

The main constraint on further studies like this one is the relative paucity of $v\sin i$ measurements for cluster members. The combination of a thorough analysis of available data and new observations with state-of-the-art multi-object spectrographs will allow more clusters to have their stellar spin orientations quantified. Building up a large sample of clusters studied in this way will enable general conclusions to be drawn about the dominant processes governing star formation.

\acknowledgments

We thank Alexander de la Vega, Geza Kovacs, Susan Mullally, Jennifer van Saders, Kevin Schlaufman, David Sing, and Jamie Tayar for discussions helpful to the initiation and completion of this project. We thank the anonymous referee for helpful comments that strengthened this paper. This paper includes data collected with the TESS mission, obtained from the MAST data archive at the Space Telescope Science Institute (STScI). Funding for the TESS mission is provided by the NASA Explorer Program. STScI is operated by the Association of Universities for Research in Astronomy, Inc., under NASA contract NAS 5–26555. This work was supported in part by a \tess\ contract (NNG14FC03C) to STScI and also a sabbatical of P.R.M. authorized by the STScI Director and hosted by JHU's Physics and Astronomy Department.

This work has made use of data from the European Space Agency (ESA) mission
{\it Gaia} (\url{https://www.cosmos.esa.int/gaia}), processed by the {\it Gaia}
Data Processing and Analysis Consortium (DPAC,
\url{https://www.cosmos.esa.int/web/gaia/dpac/consortium}). Funding for the DPAC
has been provided by national institutions, in particular the institutions
participating in the {\it Gaia} Multilateral Agreement.

This work includes data products that have been processed by
the Cambridge Astronomy Survey Unit (CASU) at the Institute of Astronomy,
University of Cambridge, and by the FLAMES/UVES reduction team at
INAF/Osservatorio Astrofisico di Arcetri. This work includes data from the GALAH survey, which is based on observations made at the Australian Astronomical Observatory, under programs A/2013B/13, A/2014A/25, A/2015A/19, and A/2017A/18. We acknowledge the traditional owners of the land on which the AAT stands, the Gamilaraay people, and pay our respects to elders past and present.

This publication makes use of data products from the Two Micron All Sky Survey, which is a joint project of the University of Massachusetts and the Infrared Processing and Analysis Center/California Institute of Technology, funded by the National Aeronautics and Space Administration and the National Science Foundation. This publication makes use of data products from the Wide-field Infrared Survey Explorer, which is a joint project of the University of California, Los Angeles, and the Jet Propulsion Laboratory/California Institute of Technology, and NEOWISE, which is a project of the Jet Propulsion Laboratory/California Institute of Technology. WISE and NEOWISE are funded by NASA. This research has made use of the AAVSO Photometric All-Sky Survey (APASS), funded by the Robert Martin Ayers Sciences Fund.

This research has made use of the SIMBAD database, operated at CDS, Strasbourg, France. This research has made use of the VizieR catalog access tool, CDS, Strasbourg, France. The original description of the VizieR service was published in A\&AS 143, 23. This work made use of NASA’s Astrophysics Data System Bibliographic Services.

\facilities{\tess\ \citep{ricker2015}, \gaia\ \citep{gaia2016,gaia2018}, VLT:Gaia-ESO Survey \citep{gilmore2012}, AAT:GALAH Survey \citep{desilva2015,buder2018}, CTIO:2MASS \citep{skrutskie2006}, WISE \citep{wise_wright2010,cutri2013}, {Hipparcos}:Tycho-2 \citep{perryman1997,hoeg2000}, AAVSO:APASS \citep{henden2016}}

\software{\texttt{astropy} \citep{astropy2018},
\texttt{astroquery} \citep{ginsburg2019},
\texttt{corner} \citep{corner},
\texttt{eleanor} \citep{feinstein2019},
        \texttt{emcee} \citep{foremanmackey2013}, \texttt{EXOFASTv2} \citep{eastman2019},
        \texttt{iPython} \citep{ipython},
        \texttt{lightkurve} \citep{lightkurve2018},
        \texttt{matplotlib} \citep{matplotlib},
        \texttt{numpy} \citep{numpy1,numpy2},
        \texttt{pandas} \citep{pandas}
        \texttt{scipy} \citep{scipy},
        \texttt{statsmodels} \citep{statsmodels},
        \texttt{TESScut} \citep{tesscut},
        \texttt{uncertainties} \citep{uncertainties}
          }

\bibliography{sample63}{}
\bibliographystyle{aasjournal}

\listofchanges
\end{document}